\documentclass[bibyear]{aa}  
\usepackage{graphicx,url}
\usepackage[varg]{txfonts}     
\usepackage{textcomp, gensymb} 
\usepackage{hyperref}                
\usepackage{pdfcomment,acronym}      
\hypersetup{
  colorlinks=true,   
  urlcolor=blue,     
  linkcolor=red,     
}
\usepackage{natbib,twoopt}
\bibpunct{(}{)}{;}{a}{}{,}    

\makeatletter
\newcommand{\bibnote}[2]{\@namedef{#1note}{#2}}
\newcommand{\biblink}[2]{\@namedef{#1link}{#2}}
\makeatother

\makeatletter
\newcommandtwoopt{\citeads}[3][][]{\href{http://adsabs.harvard.edu/abs/#3}%
{\def\hyper@linkstart##1##2{}%
\let\hyper@linkend\@empty\citealp[#1][#2]{#3}}\biblink{#3}{\href{http://adsabs.harvard.edu/abs/#3}{ADS}}}
\newcommandtwoopt{\citepads}[3][][]{\href{http://adsabs.harvard.edu/abs/#3}%
{\def\hyper@linkstart##1##2{}%
\let\hyper@linkend\@empty\citep[#1][#2]{#3}}\biblink{#3}{\href{http://adsabs.harvard.edu/abs/#3}{ADS}}}
\newcommandtwoopt{\citetads}[3][][]{\href{http://adsabs.harvard.edu/abs/#3}%
{\def\hyper@linkstart##1##2{}%
\let\hyper@linkend\@empty\citet[#1][#2]{#3}}\biblink{#3}{\href{http://adsabs.harvard.edu/abs/#3}{ADS}}}
\newcommandtwoopt{\citeyearads}[3][][]%
{\href{http://adsabs.harvard.edu/abs/#3}
{\def\hyper@linkstart##1##2{}%
\let\hyper@linkend\@empty\citeyear[#1][#2]{#3}}\biblink{#3}{\href{http://adsabs.harvard.edu/abs/#3}{ADS}}}
\makeatother

\newacro{ADS}{Astrophysics Data System}
\newacro{NLTE}{non-local thermodynamic equilibrium}
\newacro{NASA}{National Aeronautics and Space Administration}

\begin{document}


\title{Temperature and Density Structure of a Recurring Active Region Jet}
\subtitle{}

\author{
Sargam M. Mulay
\and 
Giulio Del Zanna
\and
Helen Mason
}

\institute{DAMTP, Centre for Mathematical Sciences, University of Cambridge, Wilberforce Road, Cambridge, CB3 0WA, UK\\
\email{smm96@cam.ac.uk}}

\date{Received ; accepted }

 \abstract{We present a study of a recurring jet observed on October 31, 2011 by SDO/AIA, Hinode/XRT and Hinode/EIS. We discuss the physical parameters of the jet such as density, differential emission measure, peak temperature, velocity and filling factor obtained using imaging and spectroscopic observations. A differential emission measure (DEM) analysis was performed at the region of the jet-spire and the footpoint using EIS observations and also by combining AIA and XRT observations. The DEM curves were used to create synthetic spectra with the CHIANTI atomic database. The plasma along the line-of-sight in the jet-spire and jet-footpoint was found to be peak at 2.0 MK. We calculated electron densities using the \mbox{Fe \textrm{XII}} ($\lambda$186/$\lambda$195) line ratio in the region of the spire (\mbox{N$_{\textrm{e}}$ = 7.6$\times$10$^{10}$ cm$^{-3}$}) and the footpoint (\mbox{1.1$\times$10$^{11}$ cm$^{-3}$}). The plane-of-sky velocity of the jet is found to be \mbox{524 km/s}. The resulting EIS DEM values are in good agreement with those obtained from AIA-XRT. There is no indication of high temperatures, such as emission from \mbox{Fe \textrm{XVII} ($\lambda$254.87)} \mbox{(log T [K] = 6.75)} seen in the jet-spire. In case of the jet-footpoint, synthetic spectra predict weak contributions from \mbox{Ca \textrm{XVII} ($\lambda$192.85)} and \mbox{Fe \textrm{XVII} ($\lambda$254.87)}. With further investigation, we confirmed emission from the \mbox{Fe \textrm{XVIII} ($\lambda$93.932~{\AA})} line in the AIA 94 {\AA} channel in the region of the footpoint. We also found good agreement between the estimated and predicted \mbox{Fe \textrm{XVIII}} count rates. A study of the temporal evolution of the jet-footpoint and the presence of high-temperature emission from the \mbox{Fe \textrm{XVIII}} (\mbox{log T [K] = 6.85}) line leads us to conclude that the hot component in the jet-footpoint was present initially that the jet had cooled down by the time EIS observed it. }



\maketitle


\section{Introduction}   \label{section1}

Solar jets are small-scale ubiquitous transients observed as collimated flows of plasma in and at the boundary of the coronal holes (coronal hole (CH) jets) and also at the edge of active regions (active region (AR) jets). Active region jets have been observed in H$\alpha$, extreme-ultraviolet (EUV) (c.f. \citeads{2011A&A...531L..13I}, \citeads{2014A&A...561A.134Z}, \citeads{2014A&A...567A..11Z},
\citeads{2016arXiv160200151M}) and X-ray wavelengths (c.f. \citeads{2008A&A...481L..57C}, \citeads{2008A&A...491..279C}) using ground-based and space-based instruments.

It has been observed that, AR jets are often associated with nonthermal type III radio bursts (\citeads{1995ApJ...447L.135K}, \citeads{2011A&A...531L..13I}, \citeads{2015MNRAS.446.3741C}, \citeads{2016arXiv160200151M}). The energetic particles follow the field lines that are open to the heliosphere and can produce impulsive, electron/ $^{3}$He rich solar energetic particle (SEP) events in the interplanetary medium (\citeads{2015ApJ...806..235N}, \citeads{2016arXiv160303258I}). Therefore, AR jets and their associated phenomena are one of the important features involved in space-weather studies.

Using imaging and spectroscopic observations at a number of wavelengths, we can probe different layers of the solar atmosphere and study the temperature structure of jets in detail. Until recently, it was difficult to carry out a detailed study of jets because of the limited spatial and temporal resolution of early instruments. The high spatial and temporal resolution of the \textit{Atmospheric Imaging Assembly} \citepads[AIA; ][]{2012SoPh..275...17L} on the Solar Dynamic Observatory (SDO) and the \textit{X-ray Telescope} (\citeads[XRT; ][]{2007SoPh..243...63G}) imaging instrument on Hinode, together with the spectral capabilities of \textit{EUV Imaging Spectrometer} (\citeads[EIS; ][]{2007SoPh..243...19C}) on Hinode have enabled us to carry out an in-depth analysis.  

A number of authors have studied the physical parameters such as velocities, density, size, location and direction of AR jets using EUV imaging observations (\citeads{2011A&A...531L..13I}, \citeads{2013ApJ...763...24K}, \citeads{2013ApJ...769...96C}, \citeads{2016arXiv160200151M}). There are also a few results available from spectroscopic observations (\citeads{2007PASJ...59S.763K}, \citeads{2008A&A...481L..57C}, \citeads{2011RAA....11.1229Y}, \citeads{2011A&A...526A..19M}, \citeads{2012ApJ...759...15M}, \citeads{2013ApJ...766....1L}). \citetads{2007PASJ...59S.763K}, \citetads{2008A&A...481L..57C} and \citetads{2011RAA....11.1229Y} studied AR jets using simultaneous EIS and XRT observations. They found that EUV and SXR jets had similar projected speeds, lifetimes and sizes and they also observed that EUV jets had the same location, direction and collimated shape as the SXR jets.

Using spectroscopic observations from EIS, \citetads{2007PASJ...59S.763K} observed a jet-footpoint; \citetads{2012ApJ...759...15M} and \citetads{2013ApJ...766....1L} observed a jet for the temperature range from \mbox{log \textit{T} [K] = 4.9 to 6.3} in their individual study of AR jets. \citetads{2011RAA....11.1229Y} reported the jet plasma temperature ranges from \mbox{log \textit{T} [K] = 4.7 to 6.3} i.e. from 0.05 to 2.0 MK and maximum electron densities ranging from \mbox{\textit{N$_{e}$} = 6.6$\times$10$^{9}$} to \mbox{\textit{N$_{e}$} = 3.4$\times$10$^{10}$ cm$^{-3}$}; whereas \citetads{2008A&A...481L..57C} reported a jet temperature ranges from \mbox{log \textit{T} [K] = 5.4} to \mbox{log \textit{T} [K] = 6.4} and found density above \mbox{log \textit{N$_{e}$} = 11 cm$^{-3}$}.

The temperature distribution of AR jets has been studied using Differential Emission Measure (DEM) methods assuming multi-thermal plasma along the line-of-sight using EUV imaging observations. \citetads{2013ApJ...763...24K} studied an AR surge using the DEM method of \citetads{2013SoPh..283....5A} and reported an average temperature of \mbox{2 MK} and a density of \mbox{4.1$\times$10$^{9}$ cm$^{-3}$}; whereas using the same DEM method, \citeads{2013ApJ...769...96C} studied another AR jet and reported a high temperature (7 MK) in the footpoint region. A recent study of twenty AR jets by \citetads{2016arXiv160200151M} using multiwavelength AIA observations reported the temperature of the jet-spire ranging from \mbox{log \textit{T} [K] = 6.2 to 6.3} and the electron density ranges from 8.6$\times$10$^{9}$ to 1.3$\times$10$^{10}$ cm$^{-3}$. They also investigated the temperature structure at the region of the footpoint, which was found to peak at \mbox{log \textit{T} [K] = 6.5} with electron number density ranging from \mbox{8.4$\times$10$^{9}$} to \mbox{1.1$\times$10$^{10}$ cm$^{-3}$}.

All the above studies provided various physical parameters of AR jets, but a detailed investigation of the temperature structure in the region of the spire and the footpoint of AR jets using simultaneous imaging and spectroscopic observations has remained elusive.

Even with existing instruments, it has been a challenge to find simultaneous imaging and spectroscopic observations of an active region jet. After a careful search through available datasets, we have found suitable observations of a recurrent jet originating from the periphery of an active region. To the best of our knowledge, this is the first comprehensive investigation of the temperature structure of the ‘jet-spire’ and the ‘jet-footpoint’ of an AR recurrent jet using simultaneous imaging and spectroscopic observations. In this study, we focus on two instances of a recurrent jet where we individually observed the ‘spire’ and the ‘footpoint’ of a jet.

In \mbox{section \ref{section2}}, we present our observations and describe the instruments from which the data has been taken for this study. \mbox{Section \ref{section3}} describes the DEM analysis techniques used to investigate the temperature structure of the jet spire and the footpoint. We also discuss the temporal evolution of the footpoint region and the change in temperature during that period. In \mbox{section \ref{section4}}, we discuss and summarise our results.

\section{Observations}  \label{section2}

In this section, we discuss observations of a recurrent active region jet made using SDO/AIA, Hinode/XRT and Hinode/EIS instruments. We describe below the instrument specifications and data processing techniques. 

\begin{figure}[!hbtp]
\begin{center}
\includegraphics[trim=3.5cm 0.3cm 4.2cm 1.2cm,width=0.4\textwidth]{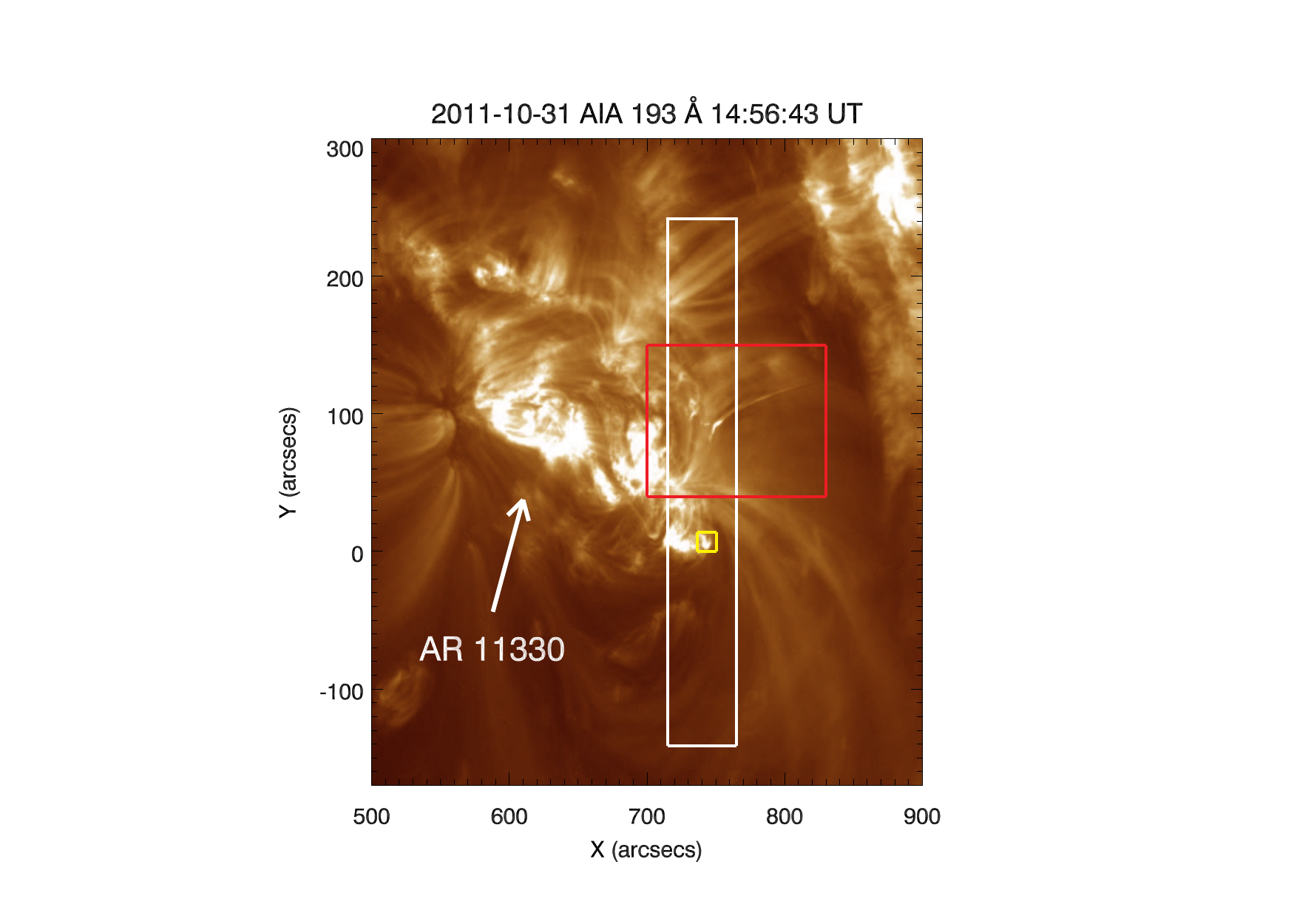}  
\caption{AIA 193~{\AA} image of an active region NOAA 11330
  \mbox{(N08 W49)} (shown by white arrow) observed on \mbox{October
    31, 2011}. The white box shows EIS raster field-of-view and the
  red box displays the field-of-view shown in fig~\ref{fig2}. The yellow box shows a moss region. \label{fig1}}
\end{center}
\end{figure}


\subsection{Atmospheric Imaging Assembly (AIA)}  \label{section2.1}

We studied a recurrent jet observed on \mbox{October 31, 2011} originating from the western edge of an active region \mbox{NOAA 11330} \mbox{(N08 W49)}. AIA observed a series of recurrent jets from \mbox{13:00 to 18:00 UT} (see online movie1.mp4) in all EUV/UV channels with its high spatial (about $\sim$ 1.2\arcsec resolution, 0.6\arcsec per pixel) and temporal (12 sec) resolution. We used six EUV wavelength channels (in brackets we indicate the spectral lines dominating each channel) : 94~{\AA} (\mbox{Fe \textrm{X}}, \mbox{Fe \textrm{XIV}}, \mbox{Fe \textrm{XVIII}}), 131~{\AA} (\mbox{Fe \textrm{VIII}}, \mbox{Fe \textrm{XXI}}), 171~{\AA} (\mbox{Fe \textrm{IX}}), 193~{\AA} (\mbox{Fe \textrm{XII}}, \mbox{Ca \textrm{XVII}}, \mbox{Fe \textrm{XXIV}}), 211~{\AA} (\mbox{Fe \textrm{XIV}}) and 335~{\AA} (\mbox{Fe \textrm{XVI}}), which are sensitive to a range of coronal temperatures (see \citeads{2010A&A...521A..21O}, \citeads{2011A&A...535A..46D}, \citeads{2013A&A...558A..73D} for a detailed description of the EUV filters in the AIA and \citeads{2015A&A...582A..56D} for their temperature responses).  Full disk AIA level 1.0 data were downloaded and processed to level 1.5 using the standard AIA software package (\textit{aia\_prep.pro}) available in the SolarSoft (\citeads[SSW ; ][]{1998SoPh..182..497F}) libraries. Further, the data were normalized by the exposure time. We carefully visually co-aligned images from different EUV channels by comparing the solar features, such as a moss region and bright patches.

\subsection{X-ray Telescope (XRT)} \label{section2.2}

These jets were also observed in X-ray wavelength channels - Ti-poly (see online movie2.mp4) and Be-thin of the XRT instrument onboard the \textit{Hinode} satellite (\citeads{2007SoPh..243....3K}). We obtained high resolution (1.02$\arcsec$) images from Ti-poly and Be-thin filters with a temporal resolution of \mbox{1 min} and \mbox{$\sim$15 min}, respectively. The XRT data were downloaded and processed using the standard XRT software package (\textit{xrt\_prep.pro}; \citeads{2014SoPh..289.2781K}) available in the SSW libraries. The processing includes the subtraction of a model dark frame, correction for vignetting and removal of high-frequency pattern noise. Further, we normalised the X-ray images by their exposure times. We employed an updated XRT filter calibration (\citeads{2011SoPh..269..169N}) in this analysis. 

\begin{figure*}[!btp]
\begin{center}
\includegraphics[trim=0.7cm 0.3cm 4cm 5cm,width=0.9\textwidth]{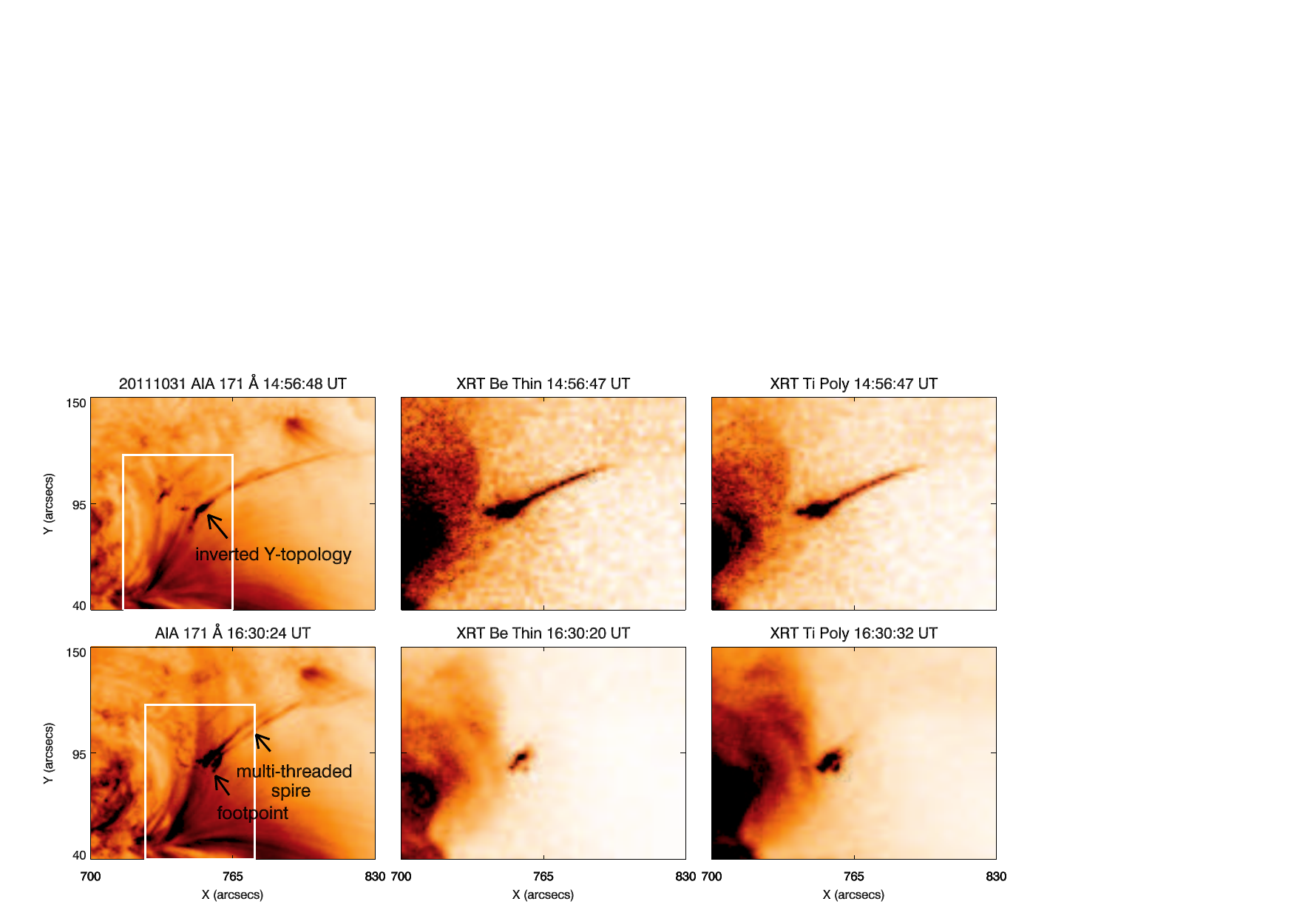} 
\caption{A recurrent jet observed on \mbox{October 31, 2011} originating from the western edge of an active region \mbox{NOAA 11330} \mbox{(N08 W49)}. The jet at \mbox{14:56 UT} (top panel) and at \mbox{16:30 UT} (bottom panel) in the AIA 171~{\AA} channel (left panel) of the AIA instrument and in the Be-thin (middle panel) and Ti-poly filters (right panel) of the XRT instrument (reverse colour images). The overplotted white boxes on the AIA 171~{\AA} images show the EIS field-of-view used in figure \ref{fig3}, \ref{fig6} and \ref{fig7}. (see online movie1.mp4 and movie2.mp4) \label{fig2}}
\end{center}
\end{figure*}

\subsection{EUV imaging spectrometer (EIS)}  \label{section2.3}

The \textit{EUV Imaging Spectrometer} (\citeads[EIS; ][]{2007SoPh..243...19C}) onboard \textit{Hinode} provides high resolution spectra in two wavelength bands, 170-211~{\AA} (SW : short wavelength) and 246-292~{\AA} (LW : long wavelength). The EIS study ‘YK\_AR\_50x384’ was run from \mbox{14:00 UT} to \mbox{17:42 UT} on October 31, 2011. The raster covered a \mbox{50 $\times$ 384 arcsec$^{2}$} area on the Sun in 18 minutes. For each raster, the \mbox{1\arcsec} slit scanned 51 positions from \mbox{west-to-east} (in the opposite direction of solar rotation) with exposure times of \mbox{20 sec}. This study includes 8 spectral windows. The EIS raster observed the ‘jet-spire’ during its fourth raster scan (\mbox{14:56:02 - 15:14:05 UT}) and the ‘jet-footpoint’ during its ninth raster scan (\mbox{16:28:17 - 16:46:20 UT}). Tables \mbox{\ref{table1} and \ref{table2}} provide details of the EIS observation and observed spectral lines. 

Figure \ref{fig1} shows the AIA 193~{\AA} image of the active region NOAA 11330 (shown by white arrow) observed on \mbox{October 31, 2011} at 14:56 UT. The white box shows the EIS raster field-of-view and the red box displays the field-of-view shown in fig~\ref{fig2}. A moss region, shown by yellow box was used to verify our method of analysis (see Appendix).


 \begin{table}[!hbtp]
 \renewcommand\thetable{1} 
 \centering
 \caption{EIS observation details}
 \resizebox{9cm}{!} {
 \begin{tabular}{lccc}

 \hline
 \hline
 &			&				 &		  			\\ 

 &{EIS details}	        &{jet-spire} 			 &{jet-footpoint}			\\ 

 &			&				 &		  			\\ 

 \hline		
 &			& 				 &		  			\\ 
				  
 &Raster start time (UT)&14:56:02   			 &16:28:17  	  			\\ 

 &Raster end time (UT)	&15:14:05 		 	 &16:46:20	  			\\ 

 &Slit width 		&1$\arcsec$			 &1$\arcsec$	  			\\
 
 &Field of view		&50$\arcsec$ $\times$ 384$\arcsec$ &50$\arcsec$ $\times$ 384$\arcsec$ 	\\ 

 &Raster cadence (min)  &18				 &18 					\\

 &Exposure time (sec)	&20 				 &20 					\\ 
 
 &Exposure number	&9,10				 &20,21					\\
 
 &Event time (UT)  	&14:58:55  		 	 &16:35:08 				\\
 
 &			&and 14:59:17			 &and 16:35:30			  	\\  
 
 &			&				 &		  			\\ 
 \hline
 \hline

  \end{tabular} \label{table1}
} 
\end{table}

 
The EIS data were processed using the standard processing routine (\textit{eis\_prep.pro}) provided in the SSW to obtain calibrated intensities in units of erg cm$^{-2}$ s$^{-1}$ sr$^{-1}$ {\AA}$^{-1}$ at each pixel in the data set. This routine flags saturated data and removes dark current, hot pixels and cosmic ray hits. The ‘cfit routine’ was applied to all of the lines at each pixel in the EIS raster. The offset \mbox{(18 pixels)} between the SW and LW CCD channels in the \mbox{Solar-\textit{Y}} direction was corrected. We employed the radiometric calibration by \citetads{2013A&A...555A..47D} to correct for the degradation of EIS instrument over time.


\begin{table}[!hbtp]
\renewcommand\thetable{2} 
\centering
\caption{List of the EIS observed emission lines}
\resizebox{7cm}{!} {
\begin{tabular}{lccc}

\hline
\hline
&					& 	    	&\\

&EIS observed lines			&{$\lambda$} 	&{log \textit{T}$_{max}$}\\ 

&					&{\AA}     	&[K]\\

\hline		
&					& 	    	&\\
			
&Fe \textrm{VIII} (bl) 	  		&186.605	&5.8 \\ 

&Fe \textrm{XII} (sbl)	  		&186.88		&6.2 \\

&Fe \textrm{XII} (sbl)	  		&195.119	&6.2 \\ 

&Fe \textrm{XVII}  	  		&254.87		&6.75 \\

&He \textrm{II} + Si \textrm{X}  	&256.32 	&4.7 \\

&Si \textrm{X}   			&256.4 	  	&6.2 \\ 

&S \textrm{XIII}			&256.686 	&6.5 \\ 

&Si \textrm{X} (bl)			&261.056 	&6.2 \\ 

&Mg \textrm{VI}				&268.991 	&5.7 \\ 

&Al \textrm{IX} (bl)			&284.042 	&6.1 \\ 

&Fe \textrm{XV} (bl)	        	&284.160 	&6.4 \\

&					& 	    	&\\

\hline
\hline

\end{tabular} 
}
\tablefoot{bl - blended lines, sbl - self-blend lines} \label{table2}

\end{table}

\begin{figure*}[!btp]
\begin{center}
\includegraphics[trim=2.5cm 0cm 5.5cm 3.5cm,width=0.6\textwidth]{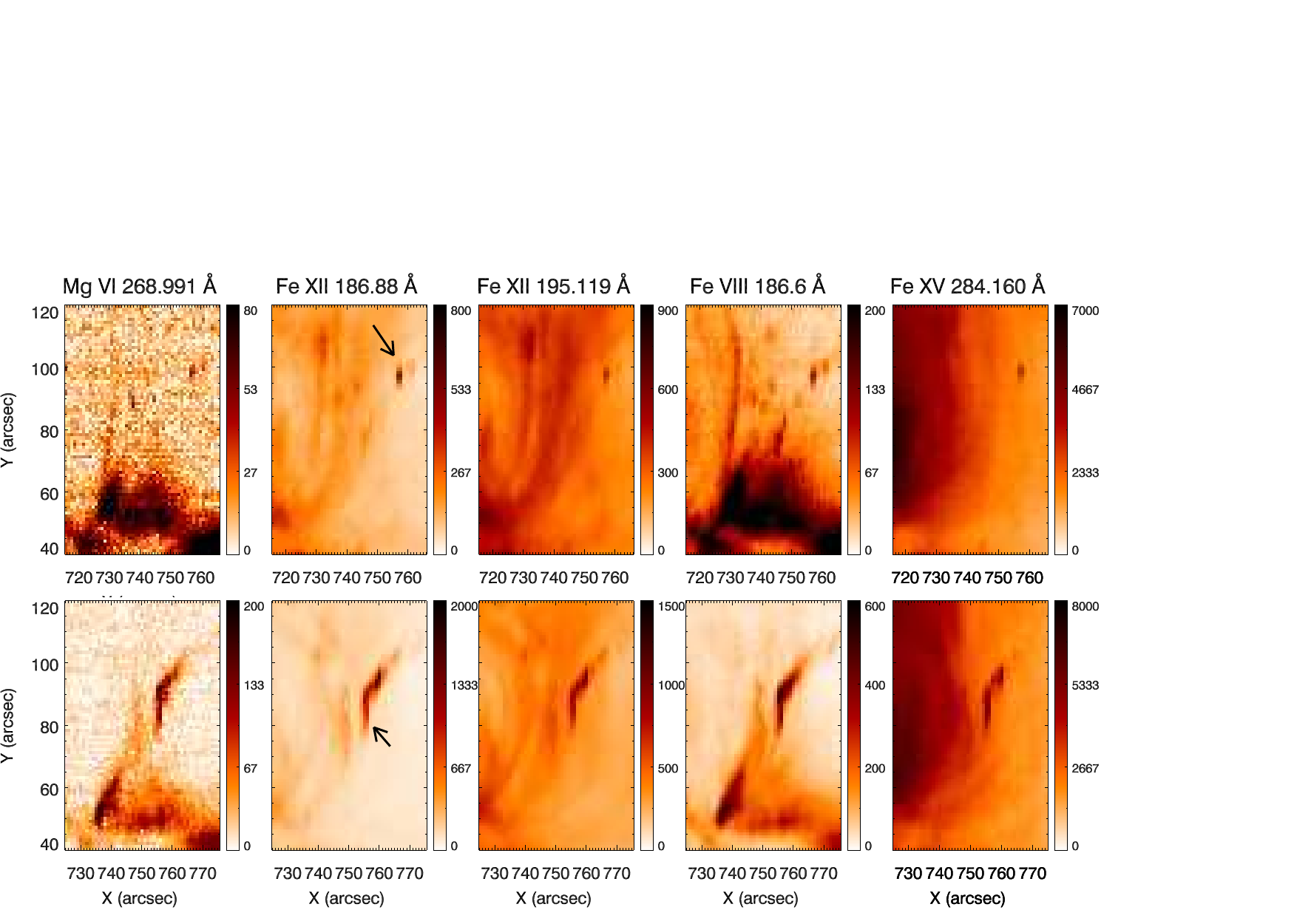}  
\caption{EIS raster images (reverse colour) showing the morphological features observed during the recurrent jet event in \mbox{Mg VI}, \mbox{Fe \textrm{VIII}}, \mbox{Fe \textrm{XII}}, and \mbox{Fe \textrm{XV}} lines. The jet-spire (observed during \mbox{14:56:02 - 15:14:05 UT}) (top panel) and the jet-footpoint (observed during \mbox{16:28:17 - 16:46:20 UT}) (bottom panel) are shown by arrows in the \mbox{Fe \textrm{XII}} images, respectively. The colourbars for the EIS rasters indicate the actual calibrated units (\mbox{{phot cm$^{-2}$ s$^{-1}$ arcsec$^{-2}$}}).  \label{fig3}}
\end{center}
\end{figure*}

\section{Data Analysis and Results}  \label{section3}

\subsection{Overview and Kinematics}  \label{section3.1}

The jet showed its recurrent nature at the same location with very similar morphological features from \mbox{13:00 UT} to \mbox{18:00 UT}. We also observed ‘plasma-blobs’ travelling along the jet spire in the AIA channels. \mbox{Figure \ref{fig2}} shows the recurrent jet (reverse colour image) at \mbox{14:56 UT} (top panel) and \mbox{16:30 UT} (bottom panel) in the \mbox{AIA 171~{\AA}} (left panel), XRT-\mbox{Be-thin} (middle panel) and XRT-\mbox{Ti-poly} (right panel) channels. The Eiffel Tower shaped jet in the \mbox{AIA 171~{\AA}} image at \mbox{14:56 UT} (shown by black arrow in the top left panel) represents an inverted Y-topology of the magnetic field lines associated with the jet footpoint structure. A multi-threaded spire is observed during its evolution which is clearly seen in the \mbox{AIA 171~{\AA}} image at \mbox{16:30 UT} (shown by black arrow in the bottom left panel). The XRT images show a similar morphology for the jet along with a bright spot at the footpoint. The overplotted white boxes on AIA 171~{\AA} images in fig. \ref{fig2} show the EIS field-of-view used in figs. \ref{fig3}, \ref{fig6} and \ref{fig7}.

The jet spire and the footpoint were observed over the range of temperatures from \mbox{log \textit{T} [K] = 4.7} to \mbox{log \textit{T} [K] = 6.4} in the EIS observations. The emission from these regions is shown by black arrows in Fig \ref{fig3} (second column). Figure \ref{fig3} shows the EIS raster images (reverse color) in calibrated units \mbox{(phot cm$^{-2}$ s$^{-1}$ arcsec$^{-2}$)} for the jet-spire (top panel) and the jet-footpoint (bottom panel) in \mbox{Mg VI ($\lambda$268.991, \mbox{log \textit{T} [K] = 5.7})}, \mbox{Fe \textrm{VIII}} ($\lambda$186.605, \mbox{log \textit{T} [K] = 5.8}), \mbox{Fe \textrm{XII} ($\lambda$186.88 and $\lambda$195.119)}, \mbox{log \textit{T} [K] = 6.2}) and \mbox{Fe \textrm{XV} ($\lambda$284.160, \mbox{log \textit{T} [K] = 6.4})} lines respectively. There is no emission observed in \mbox{Fe \textrm{XVII}} line \mbox{($\lambda$254.87)} (\mbox{log \textit{T} [K] = 6.75}) at the region of the spire and the footpoint.   

\begin{figure}[!hbtp]
\begin{center}
\includegraphics[trim=0cm 0.1cm 1.5cm 0cm,width=0.38\textwidth]{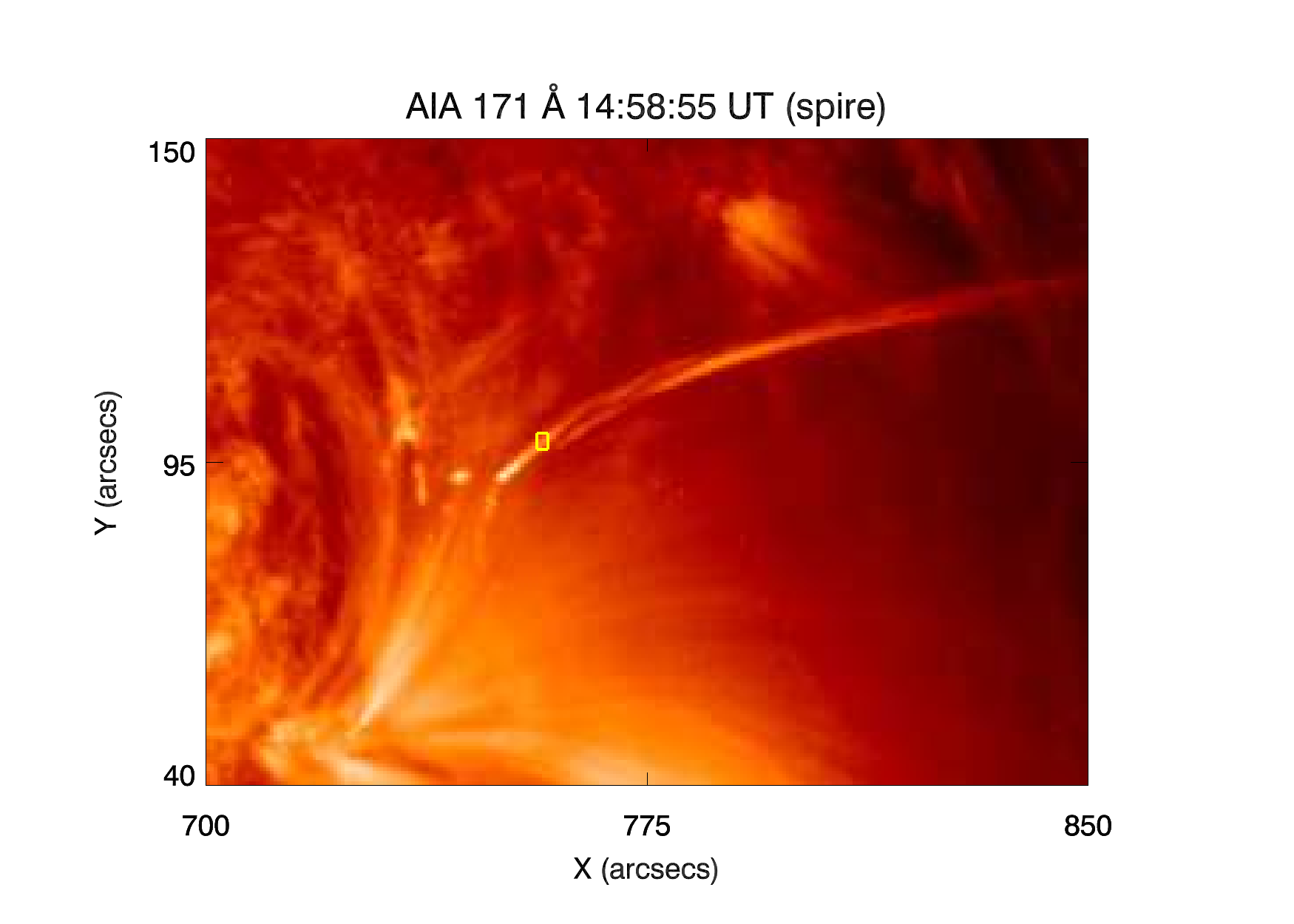}
\includegraphics[trim=0.8cm 0.1cm 0cm 0cm,width=0.48\textwidth]{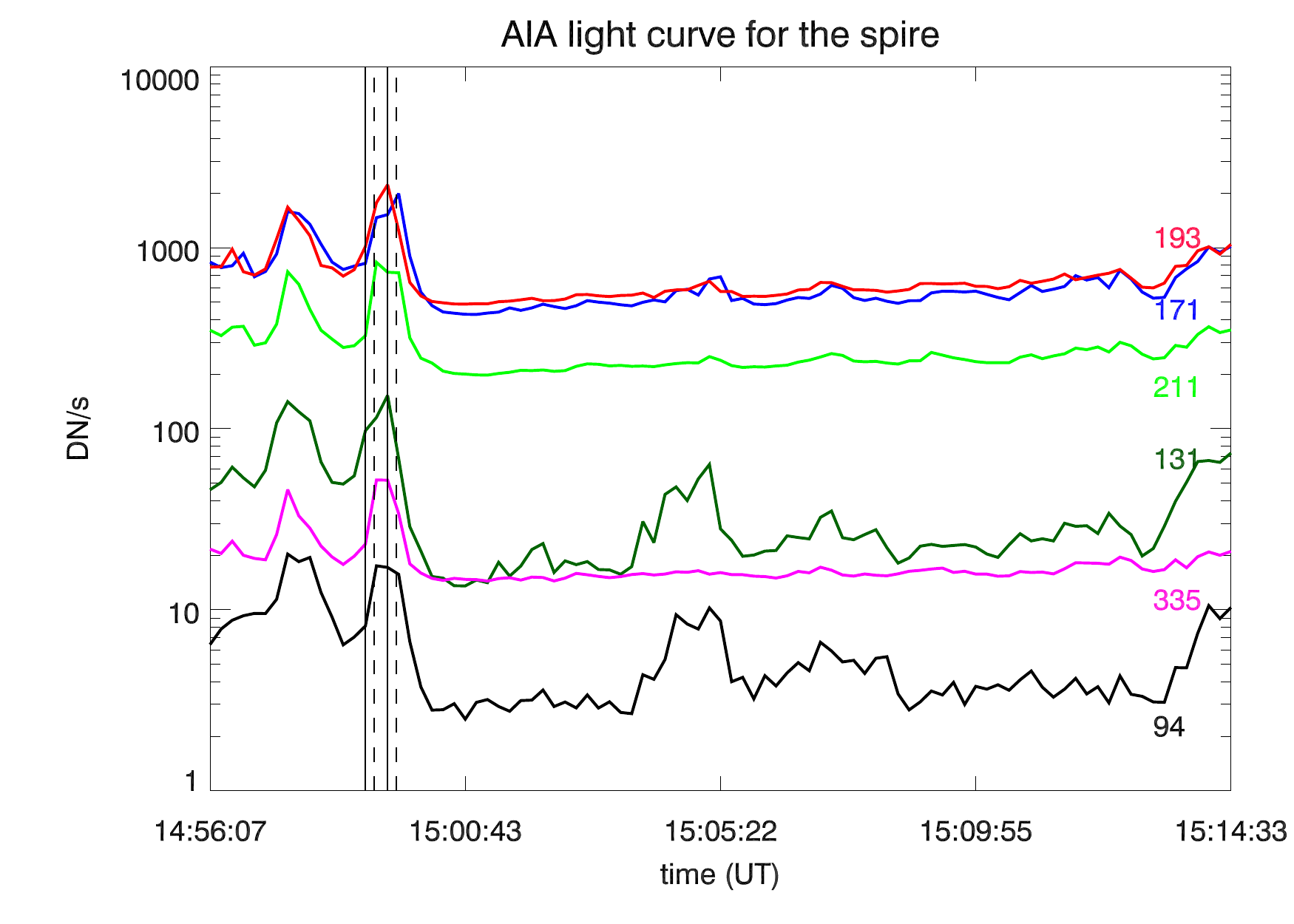}
\caption{Top panel : AIA 171~{\AA} image at 14:58 UT. The overplotted yellow box indicates the region of the jet-spire selected for further analysis shown as a small box in fig.~\ref{fig6} (top panel - first column). Bottom panel : Temporal  evolution of the jet-spire calculated in the same small box using all AIA coronal channels. The overplotted solid lines indicate the EIS slit position start time for two exposures where EIS observed the spire and dashed lines indicate the slit position end times. \label{fig4}}
\end{center}
\end{figure}

\begin{figure}[!hbtp]
\begin{center}
\includegraphics[trim=0cm 0.1cm 1.5cm 0cm,width=0.38\textwidth]{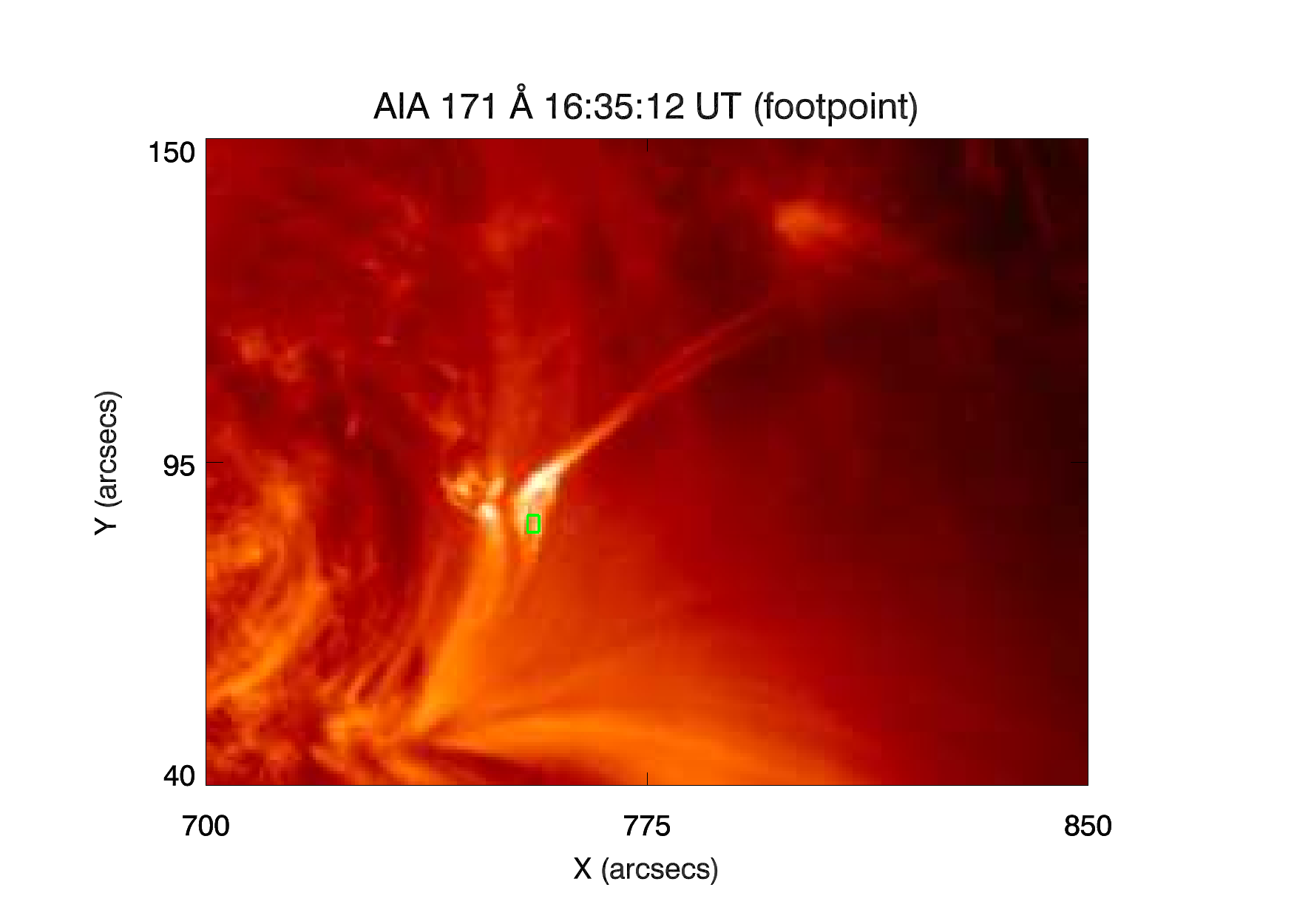}
\includegraphics[trim=0.8cm 0.1cm 0cm 0cm,width=0.48\textwidth]{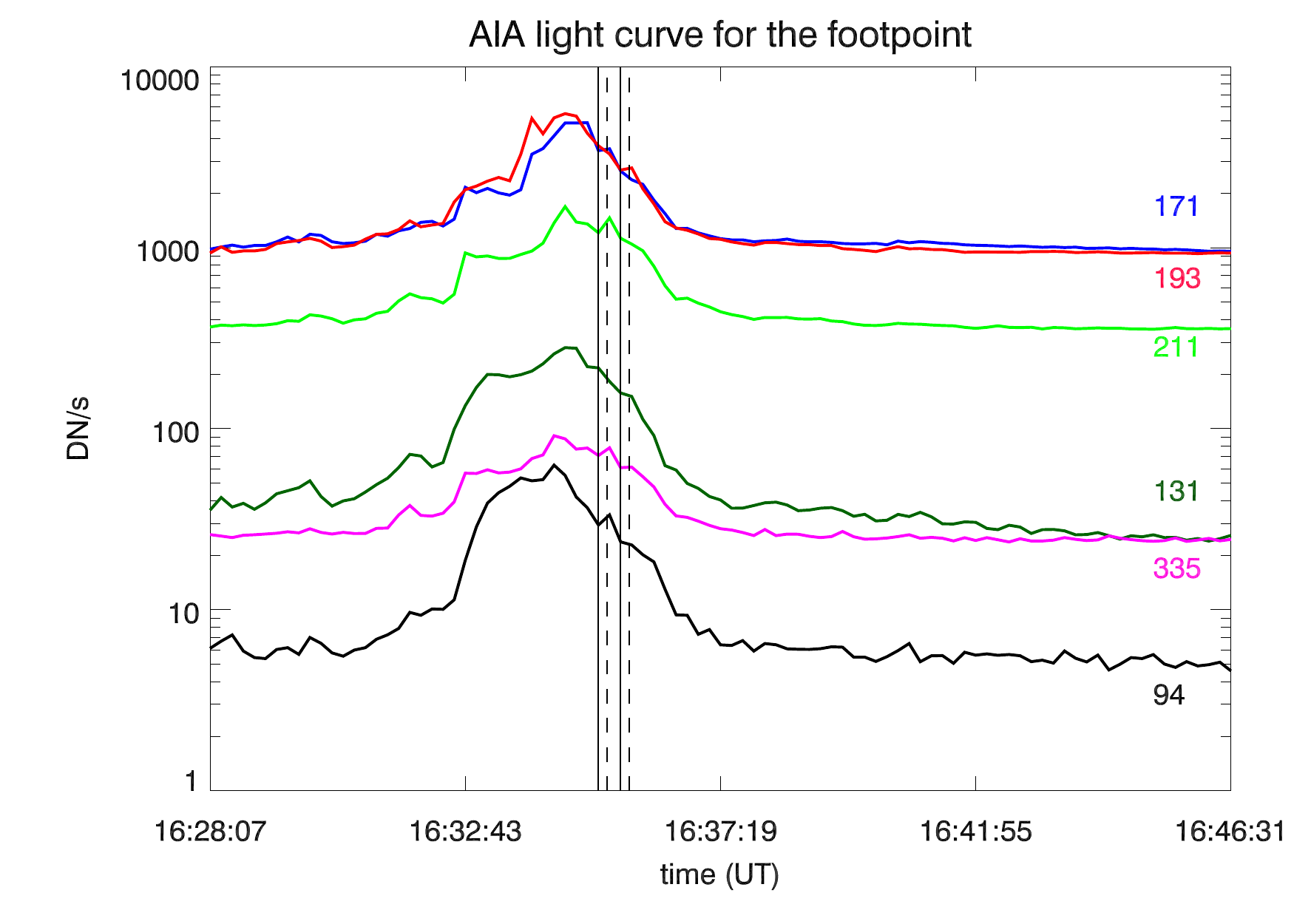}
\caption{same as fig.\ref{fig4} for the region of the jet-footpoint (shown as a small box in fig.~\ref{fig6} - bottom panel - first column). \label{fig5}}
\end{center}
\end{figure}

The highly dynamic activity of the jet was clearly observed in the AIA channels during its evolution (see online movie1.mp4). Figure~\ref{fig4} (top panel) shows the jet in the \mbox{AIA 171~{\AA}} channel at \mbox{14:58 UT} and overplotted yellow box shows the region where the EIS observed an emission from the spire. We obtained a temporal evolution (bottom panel) of this region in all AIA channels to see the variability of the jet-spire structure during its evolution. We also investigated the AIA count rates in the region of the footpoint. Figure~\ref{fig5} (top panel) shows the jet in the \mbox{AIA 171~{\AA}} channel at \mbox{15:35 UT} and the overplotted green box shows the region where EIS observed the emission from the footpoint. We also obtained the temporal evolution (bottom panel) of this region in all AIA channels and observed the variability in the jet-footpoint region. The overplotted solid lines indicate the EIS slit position start time for two exposures where EIS observed the spire and footpoint and dashed lines indicate the slit position end times. The light curves show a sudden increase in the count rates in all AIA channels during the period for the jet spire study with EIS (see fig. \ref{fig4}). For the footpoint study (see fig. \ref{fig5}), the AIA count rates at the time of EIS observations are past their peak and decreasing steadily.

\subsection{AIA-EIS-XRT co-alignment}  \label{section3.2}

A direct comparison of AIA images with EIS spectroscopic data is not straightforward for several reasons; the AIA channels are multi-thermal; the AIA and EIS pixel size and spatial resolution differs; the EIS slit scans the field-of-view to produce a raster; the exposure times for the AIA images and EIS raster scan are also different. 

The \mbox{AIA 193~{\AA}} channel, which is dominated by \mbox{Fe \textrm{XII}}, is most suitable for a direct comparison. EIS observes the \mbox{Fe \textrm{XII}} line ($\lambda$195.119, \mbox{log \textit{T} [K] = 6.2}) in the SW channel. We followed a procedure given by \citetads{2011A&A...535A..46D} to co-align the images from the \mbox{AIA 193~{\AA}} channel with the EIS \mbox{Fe XII} raster : firstly, we convolve each simultaneous \mbox{AIA 193~{\AA}} image with each EIS slit position by considering a Gaussian PSF of 2$\arcsec$ full-width-half-maximum (FWHM). The co-temporal AIA images taken during the EIS exposures are averaged and then rebinned to the ‘EIS pixel’ size (in this case, it is 1\arcsec) to obtain a slice of corresponding averaged AIA image. We then built a time-averaged ‘rebinned’ map of \mbox{AIA 193~{\AA}} channel. The data from the other AIA channels together with the XRT channels are co-aligned according to emission observed in the AIA 193~{\AA} channel. In this analysis, we have used only the high cadence (1 min) images from the XRT Ti-poly filter. We also found an offset of 12 pixels in \textit{X}-direction and 36 pixels in \textit{Y}-direction between rebinned AIA maps and EIS raster maps and we corrected all the maps.

\begin{figure*}[!btp]
\begin{center}
\includegraphics[trim=1.5cm 0cm 6cm 1.9cm,width=0.65\textwidth]{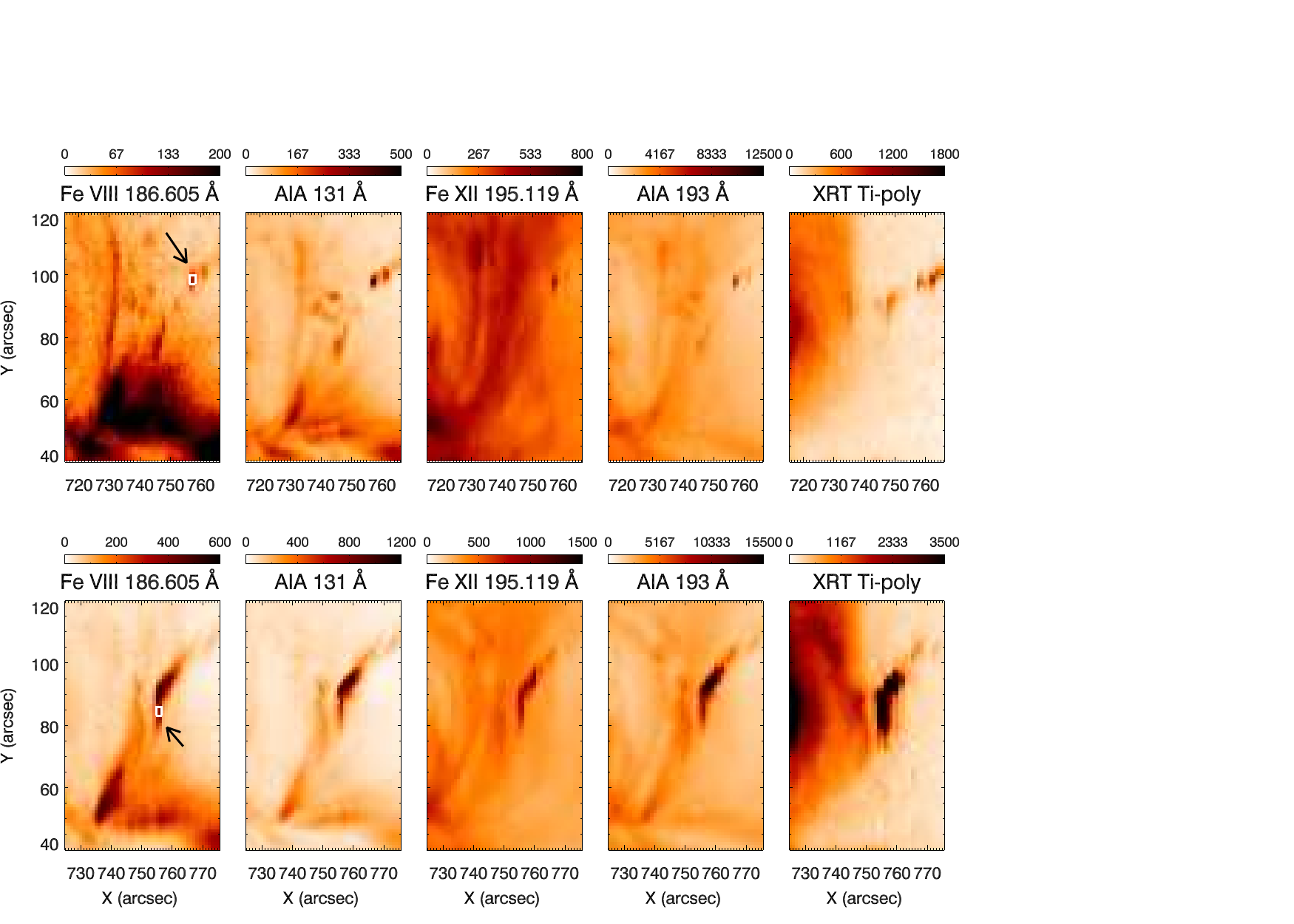} 
\caption{AIA-EIS-XRT co-aligned images (reverse colour) of the jet-spire (top panel) and the jet-footpoint (bottom panel). Column 1 and 3 : EIS intensity images in \mbox{Fe \textrm{VIII} ($\lambda$186.605)} and \mbox{Fe \textrm{XII} ($\lambda$195.119)} respectively. Column 2 and 4 : a time-averaged ‘rebinned’ maps in the \mbox{AIA 131 and 193~{\AA}} channels respectively. Column 5 : a coaligned map in the Ti-poly filter. The colourbars for the EIS rasters indicate the actual calibrated units ({phot cm$^{-2}$ s$^{-1}$ arcsec$^{-2}$}) and for AIA images, the units are DN s$^{-1}$ EIS pixel$^{-1}$. The regions shown by white boxes (indicated by black arrows) are used for the DEM analysis (see section \ref{section3.6}).
\label{fig6}}
\end{center}
\end{figure*}

Figure \ref{fig6} shows the jet-spire (top panel) and the jet-footpoint (bottom panel) observed with EIS \mbox{Fe \textrm{VIII}} \textit{(column 1)} and \mbox{Fe \textrm{XII}} \textit{(column 3)}, time-averaged rebinned maps in the \mbox{AIA 131} \textit{(column 2)} and \mbox{193~{\AA}} \mbox{\textit{(column 4)}} channels and XRT Ti-poly filter \textit{(column 5)}. The two white overplotted boxes (shown by black arrows in \textit{column 1}) indicate the region of the spire and the footpoint used for the DEM analysis (see section \ref{section3.6}). The jet-spire appears as a ‘small blob’ in the \mbox{Fe \textrm{VIII}} and \mbox{Fe \textrm{XII}} raster images. The time-averaged AIA 193~{\AA} images are in very good agreement with the emission seen in \mbox{Fe \textrm{XII}} by EIS. Also, the \mbox{AIA 131~{\AA}} image shows a similar structure and morphology for the jet as in the EIS \mbox{Fe \textrm{VIII}} observation, indicating that the emission observed in the \mbox{AIA 131~{\AA}} comes primarily from the lower temperature contribution, mainly \mbox{Fe \textrm{VIII}}.


\subsection{Spectroscopic techniques} \label{section3.3}

We used spectroscopic techniques to derive the basic physical plasma parameters such as electron density, differential emission measure and filling factor. 

The intensity of an optically thin emission line can be written as

\begin{equation}
 I = 0.83~Ab(z)~\int G(T_\textrm{e},N_\textrm{e})~N^2_\textrm{e} dh
\label{eq:int}
\end{equation}

where the factor 0.83 is the ratio of protons to free electrons, $Ab(z)$ is the elemental abundances, $N_\textrm{e}$ is the electron number density, $h$ is the column depth of the emitting plasma along the \mbox{line-of-sight}. $G(T_\textrm{e},N_\textrm{e})$ is the contribution function which contains all the atomic parameters for each spectral line and is defined as

\begin{equation}
 G(T_\textrm{e}, N_\textrm{e}) =  \frac{hc}{4\pi \lambda_{\textrm{i},\textrm{j}}} \frac{A_{\textrm{ji}}}{N_\textrm{e}} \frac{N_j (X^{\textrm{+m}})}{N(X^{\textrm{+m}})} \frac{N(X^{\textrm{+m}})}{N(X)}   
\label{eq:g}
\end{equation}

where $\textrm{i}$ and $\textrm{j}$ are the lower and upper levels, $A_{\textrm{ji}}$ is the spontaneous transition probability, $\frac{N_\textrm{j} (X^{\textrm{+m}})}{N(X^{\textrm{+m}})}$ is the population of level $\textrm{j}$ relative to the total $N(X^{\textrm{+m}})$ number density of ion $X^{\textrm{+m}}$ and is a function of electron temperature and density, $ \frac{N(X^{\textrm{+m}})}{N(X)}$ is the ratio of the ion $X^{\textrm{+m}}$ relative to the total number density of element X.

A differential emission measure DEM(T) can be defined as :

\begin{equation}
 \int_{T} DEM(T) dT = \int N_{\textrm{e}}^2 dh 
\label{eq:dem}
\end{equation}

DEM represents the amount of plasma along the \mbox{line-of-sight} that emit observed radiation and has temperature between T and T+dT. The emission measure (EM) can be calculated by integrating the DEM over a given temperature range. For an individual ion, this temperature range is about \mbox{0.3 (in log \textit{T})},

\begin{equation}
 EM = \int N_{\textrm{e}}^2 dh = \int_{T} DEM(T) dT  [cm^{-5}]
\label{eq:emd}
\end{equation}

The above equation is valid only if the volume is completely filled by the emitting plasma, otherwise a filling factor should be used. The filling factor of a plasma ${(\phi)}$ is the fraction of a volume that is contributing to the observed emission. 

\begin{equation}
 \phi=\frac{EM}{N_\textrm{e}^2 h}
\label{eq:phi}
\end{equation}
where $EM$ is measured using Equation~\ref{eq:emd} and N$_\textrm{e}$ is measured by the line ratio techniques (see section~\ref{section3.4}).

\subsection{Density measurement}  \label{section3.4}

The ratio of intensities for certain pairs of emission lines emitted from the same ion, for example some coronal ions, can be used as a density diagnostic. Due to a broad range of sensitivity of the Fe \textrm{XII} lines (\mbox{log \textit{N$_e$} = 8-12}), we used the density-sensitive pair ratio ($\lambda$186/$\lambda$195) to calculate an electron density profile in the jet. We used theoretical line intensity ratios of \mbox{Fe \textrm{XII}} ($\lambda$(186.854+186.887)/$\lambda$(195.119+195.179)) lines from the \mbox{CHIANTI} atomic database v.8 (\citeads{1997A&AS..125..149D}, \citeads{2015A&A...582A..56D}). 

\begin{figure}[!hbtp]
\begin{center}
\includegraphics[trim=0.1cm 0cm 1.7cm 1.1cm,width=0.55\textwidth]{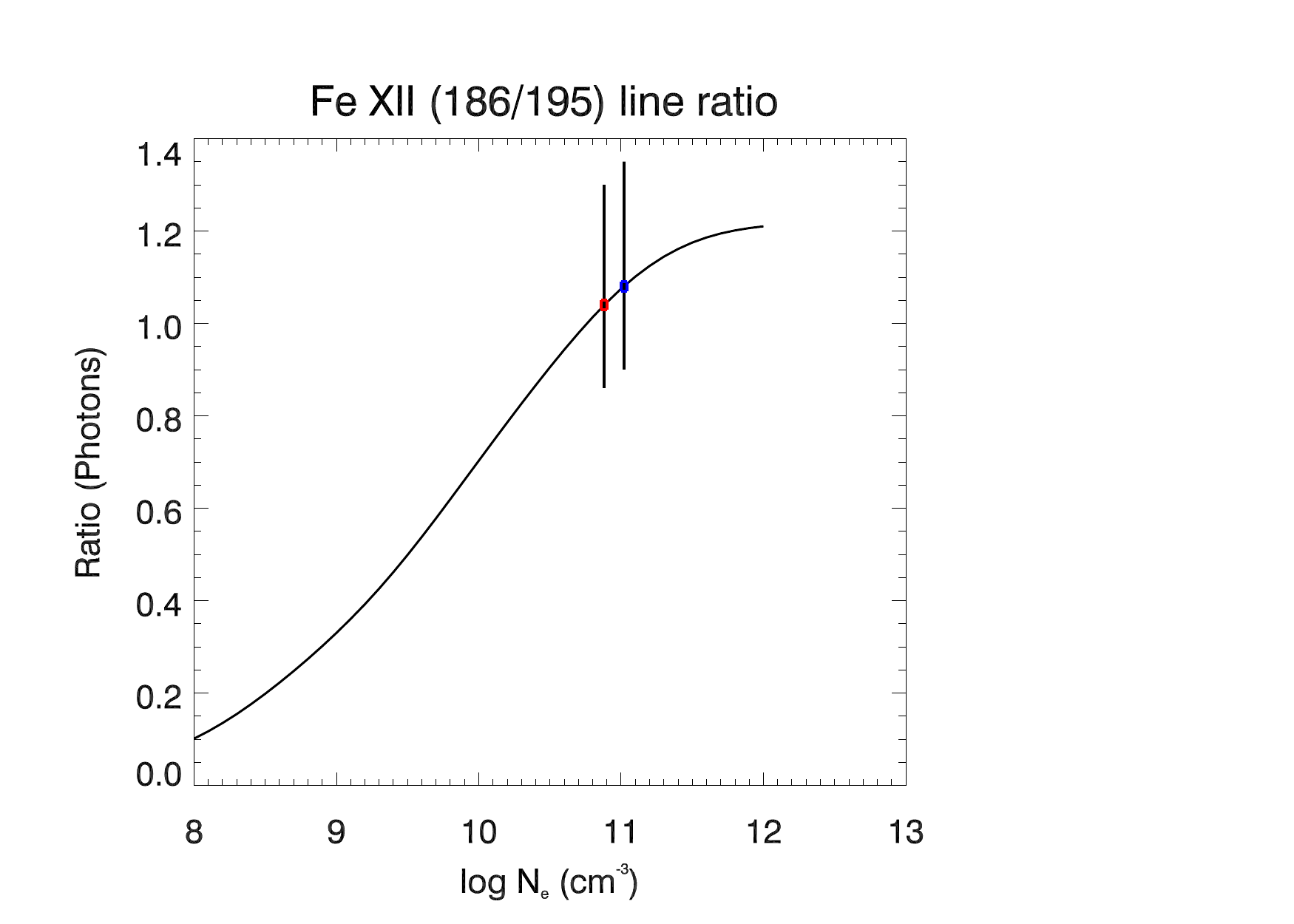}
\includegraphics[trim=0cm 0cm 1.0cm 0cm, width=0.46\textwidth]{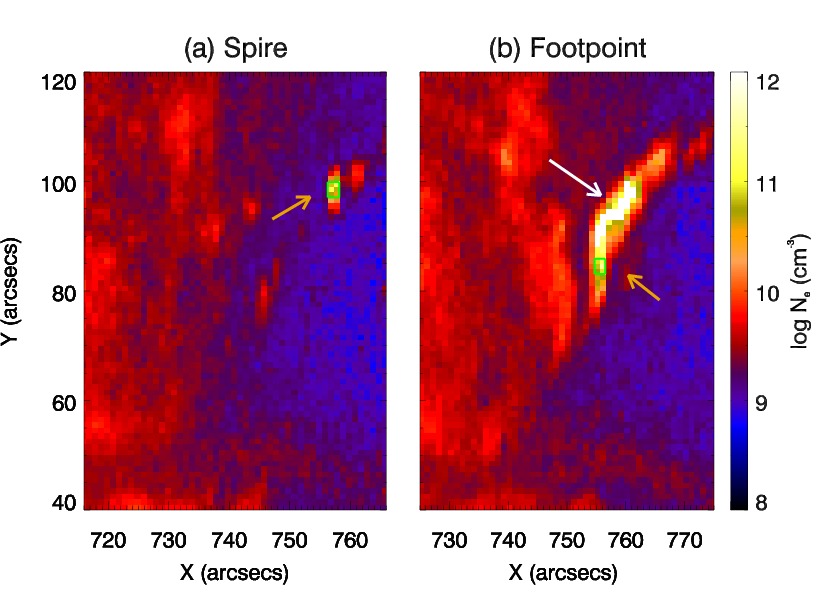}
\caption{Top panel : The theoretical intensity of the \mbox{Fe \textrm{XII}} ($\lambda$186/$\lambda$195) line ratio as a function of electron density. The density ratio for the jet-spire and jet-footpoint are shown by red and blue boxes respectively. The vertical lines overplotted on the curve shows the errors estimated by considering 20\% uncertainty. Bottom panel : electron density maps for the (a) jet-spire and the \mbox{(b) jet-footpoint}. The electron densities are in logarithmic scale in cm$^{-3}$ units. The regions used for the density diagnostics are shown by green boxes (indicated by orange arrows).   \label{fig7}}
\end{center}
\end{figure}

\mbox{Figure \ref{fig7}} (top panel) curve represents the theoretical line intensity ratio as a function of electron density. The density ratio for the jet-spire and jet-footpoint are shown by red and blue boxes respectively. The vertical lines overplotted on the curve show the errors estimated by considering 20\% uncertainty (taking account of uncertainties in the atomic data) in the calibrated intensities. The bottom panel shows \mbox{Fe \textrm{XII}} density maps for the jet-spire (a) and the jet-footpoint (b). The emission was observed for two EIS raster positions shown by orange arrows.

We calculated electron densities for the jet-spire and the jet-footpoint (at the locations given by the green boxes in fig.~\ref{fig7} (bottom panel) - shown by orange arrows) which are found to be \mbox{\textit{N}$_\textrm{e}$ = 7.6$\times$10$^{10}$} and 1.1$\times$10$^{11}$ cm$^{-3}$ respectively. We note that with 20\% error bars these values could be in the high density limit. We used these density values in the DEM analysis of the spire and the footpoint (see section \ref{section3.6}). 

From the observation we found that the $\lambda$186/$\lambda$195 pair ratio is greater than 1.2 in parts of the jet, shown by white arrow in the density map  (fig. \ref{fig7} bottom panel (b)). This indicates that the region has a high density (\mbox{log \textit{N$_e$} $\geq$ 11.5}), which is above the range of sensitivity of \mbox{Fe \textrm{XII}} lines ratio diagnostics.

\subsection{Velocity measurement}  \label{section3.5}

\begin{figure}[!hbtp]
\begin{center}
\includegraphics[trim=0cm 0.1cm 0.5cm 0.2cm,width=0.45\textwidth]{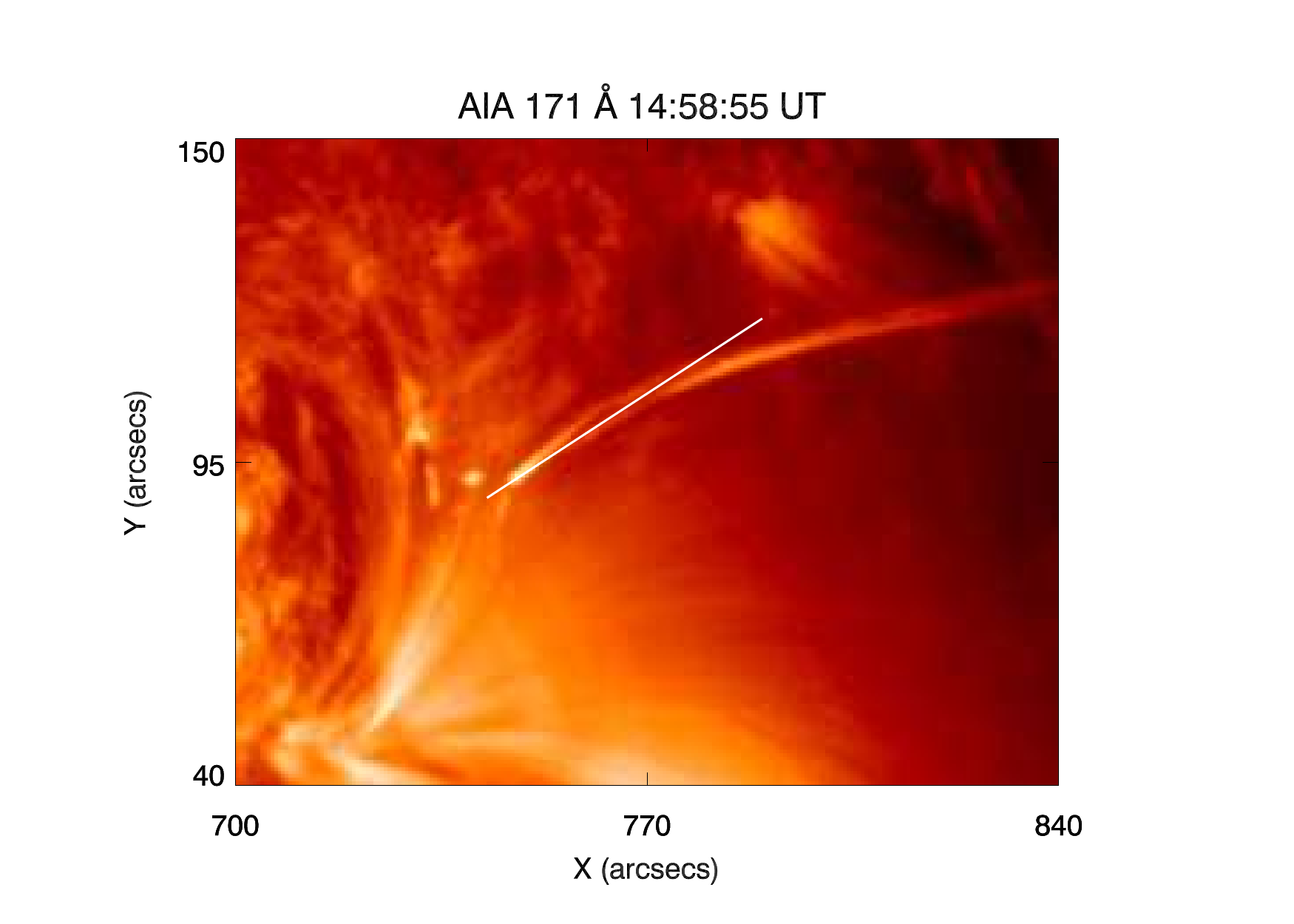}
\includegraphics[trim=1.5cm 0.1cm 0cm 0.2cm,width=0.45\textwidth]{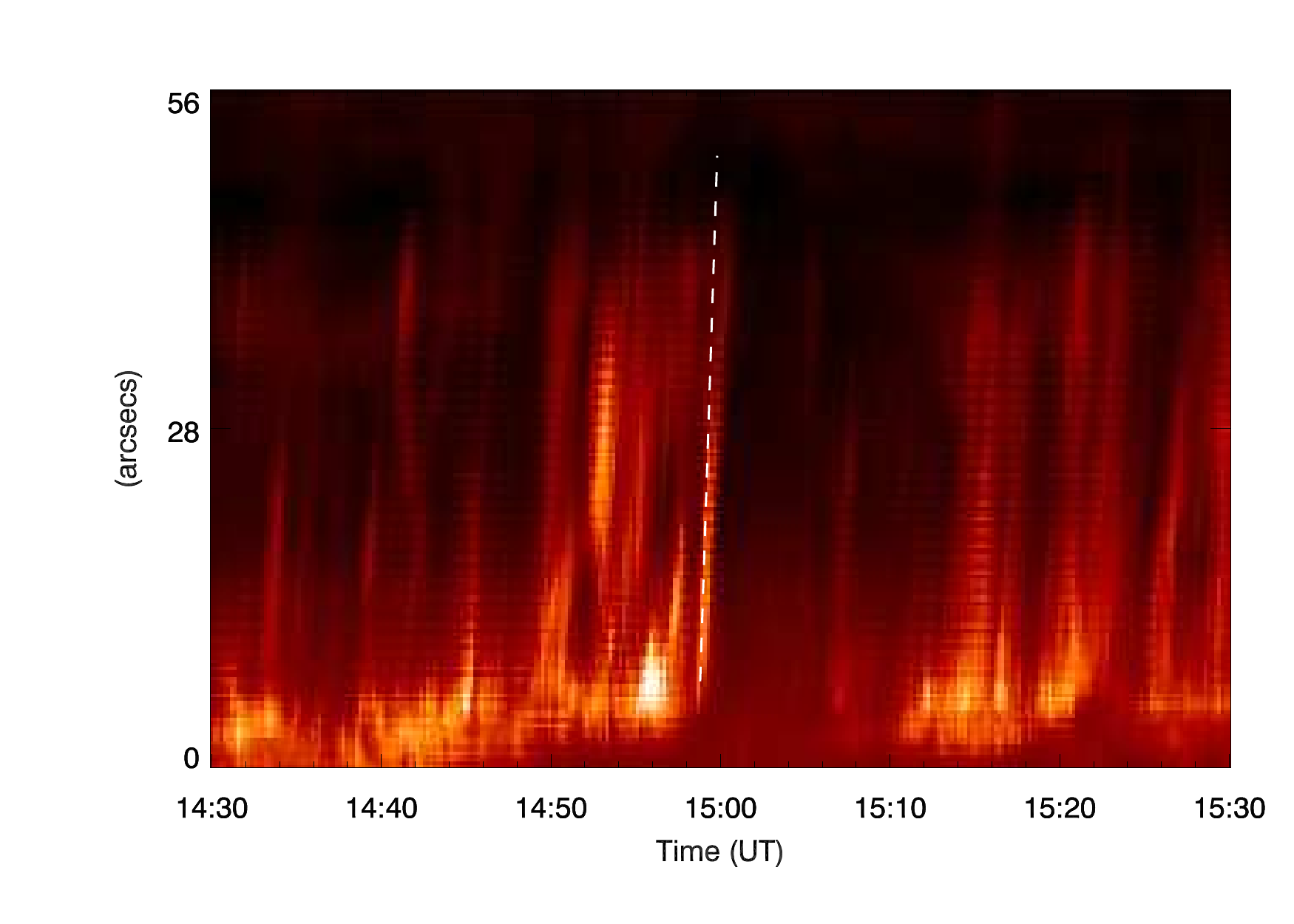}
\caption{Top panel : the jet evolution in the AIA 171~{\AA} channel at \mbox{14:58:55 UT}. The overplotted white line shows an artificial slit which is used to produce a time-distance plot. Bottom panel : time-distance plot along the jet spire for an hour from \mbox{14:30 to 15:30 UT}. The white dashed line is used for the velocity calculation which is found to be \mbox{524 km/s}.\label{fig8}}
\end{center}
\end{figure}

We performed a time-distance analysis to obtain the plane-of-sky velocity of the jet using the AIA 171~{\AA} channel images (dominated by \mbox{Fe \textrm{IX}}; \mbox{log \textit{T} [K] = 5.8}). Figure \ref{fig8} (top panel) shows the \mbox{AIA 171~{\AA}} image of the jet and the white line represents an artificial slit along the direction of the jet spire which is used to obtain the time-distance plot. The bottom panel shows the jet activity for an hour from \mbox{14:30 to 15:30 UT}. The near vertical features in the time-distance plot shows recurrent jets originating from the same location with similar morphology. We calculated a velocity for the jet at \mbox{14:58:55 UT} (the same time when EIS observed the jet-spire) and it was found to be \mbox{524 km/s} initially.

\subsection{Differential Emission Measure (DEM)}   \label{section3.6}

We have performed a DEM analysis to determine the distribution of plasma as a function of temperature. In this section, we present our results from the DEM analysis using EIS spectroscopic observations. Also, we discuss DEM results obtained by combining the AIA and the XRT imaging observations. 

\subsubsection{EIS DEM} \label{section3.6.1}

We used a set of EIS lines (see Table \ref{table3}) emitted by various elements over a wide range of temperatures (\mbox{log \textit{T} [K] = 5.7} to \mbox{log \textit{T} [K] = 6.4}) to calculate the line-of-sight DEM. Emission for the jet-spire and the jet-footpoint covered two consecutive slit positions, 1$\arcsec$ wide, during \mbox{14:58:55} - \mbox{14:59:17 UT} (exposure number - 9 and 10) and \mbox{16:35:08} - \mbox{16:35:30 UT} (exposure number 20 and 21), respectively.

\begin{figure}[!hbtp]
\begin{center}
\includegraphics[trim=2.0cm 12.0cm 0.1cm 4.0cm,width=0.52\textwidth]{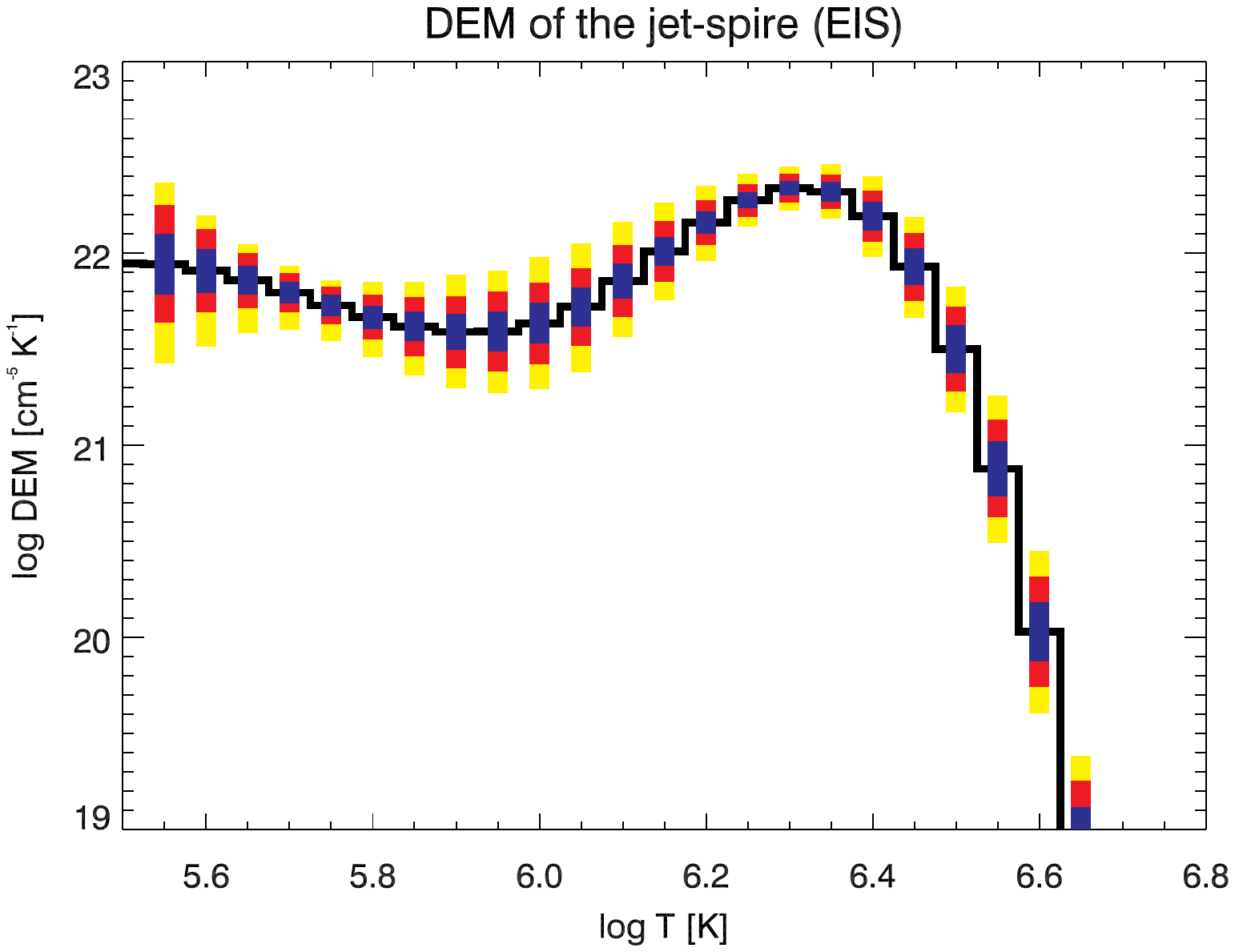} 
\includegraphics[trim=2.0cm 12.0cm 0.1cm 4.0cm,width=0.52\textwidth]{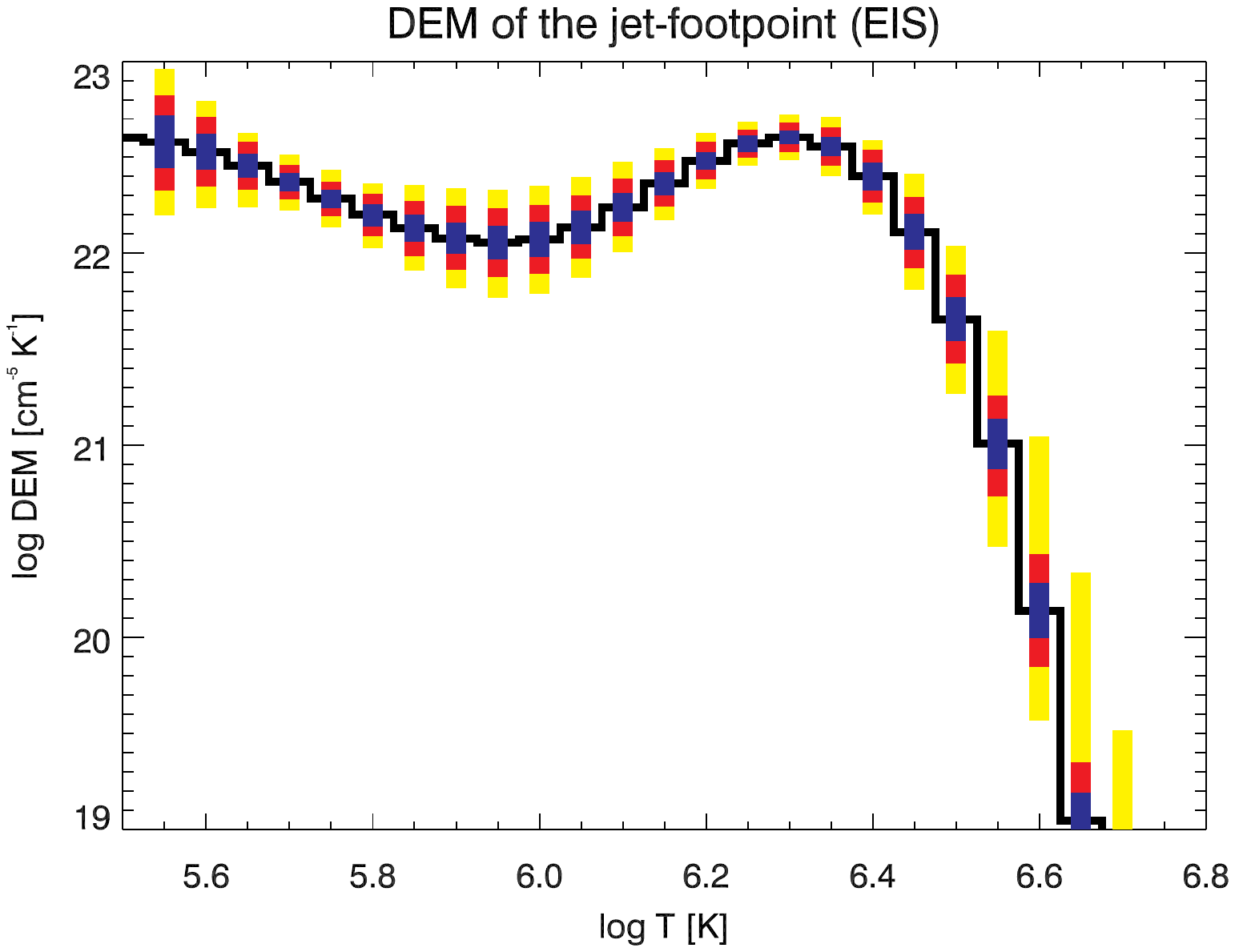}
\caption{EIS DEM curves for the jet-spire (top panel) and the jet-footpoint (bottom panel). The DEMs are obtained from the ‘XRT\_DEM\_ITERATIVE2’ method (black curve). The blue rectangles represent 50\%, red rectangles 80\% and yellow rectangles 95\% of the MC solutions in each temperature bin. \label{fig9}}
 
\end{center}
\end{figure}



\begin{table*}[!btp]
\renewcommand\thetable{3} 
 \centering
\caption{List of the EIS observed emission lines used to constrain DEM curves}
\resizebox{14cm}{!} {
\begin{tabular}{lccccccc}

\hline
\hline
&&&&\\
&EIS observed lines			&{$\lambda$} 	&{log \textit{T}$_{max}$} &{Observed}  &{Predicted}	&{Observed}	&{Predicted}	\\ 

&{used for the DEM}			&({\AA})     	&(K)			  &{(spire)}	&{(spire)}	&{(footpoint)}	&{(footpoint)}	\\

\hline		
&					& 	    	& 			  &		&		&		&	\\
			
&Fe \textrm{VIII} (bl) 	  		&186.605	&5.8 			  &445		&562 (26\%) 	&1617		&1982 (23\%)	\\ 

&Fe \textrm{XII}   	  		&195.119	&6.2                      &1707		&1755 (3\%)	&3270		&3415 (4\%)	\\ 

&Si \textrm{X}   			&261.056  	&6.2 			  &408		&298 (-27\%)	&790		&629 (-20\%)	\\ 

&Mg \textrm{VI}				&268.991 	&5.7 			  &106		&62 (-42\%)	&355		&240 (-32\%)	\\ 

&Fe \textrm{XV} (bl)			&284.160 	&6.4 			  &9221		&9943 (8\%)	&16270		&17170 (6\%)	\\

&					& 	   	& 			  &		&		&		&	\\

\hline
\hline

\end{tabular} 
}
\tablefoot{Column 4 and 5 represents observed and predicted intensities (obtained from the EIS DEM) for the jet-spire region, Column 6 and 7 represents observed and predicted intensities (obtained from the EIS DEM) for the jet- footpoint region, EIS intensities are in erg cm$^{-2}$ s$^{-1}$ sr$^{-1}$ units, bl - blended lines, percentage difference between observed and predicted intensities are give in parenthesis in columns 5 and 7.} \label{table3}

\end{table*}

We have extracted a small region of 2$\arcsec$ $\times$ 3$\arcsec$ (2$\times$3 pixels) area around the jet spire and footpoint, which are shown as small white boxes in Fe \textrm{VIII} raster image in \mbox{figure \ref{fig6}} (first column). We then obtained an EIS averaged spectrum for these regions and performed a DEM analysis on the EIS intensities (see Table \ref{table3}) using the ‘XRT\_DEM\_ITERATIVE2’ method (\citeads{2004IAUS..223..321W}), available within SSW. We used IDL routine ‘CHIANTI\_DEM’ to calculate the contribution function \mbox{C(T, $\lambda$$_{\textrm{ij}}$, N$_{\textrm{e}}$)} at a constant electron density. We used the electron density calculated from the \mbox{Fe \textrm{XII}} lines intensity ratio technique (see \mbox{section \ref{section3.4}}) and a set of ‘photospheric’ abundances by \citetads{2009ARA&A..47..481A} in this calculation.

Figure \ref{fig9} shows the DEM curves for the jet-spire (top panel) and jet-footpoint (bottom panel) obtained from the EIS data in the temperature range between \mbox{log \textit{T} [K] = 5.5} and 7.1 with an interval of log \textit{T} [K] = 0.05. The DEM for the jet-spire peaks at \mbox{log \textit{T} [K] = 6.3} with \mbox{DEM = 2.1$\times$10$^{22}$ cm$^{-5}$ K$^{-1}$}, and for the jet-footpoint, it also peaks at \mbox{log \textit{T} [K] = 6.3} with \mbox{DEM = 3.9$\times$10$^{22}$ cm$^{-5}$ K$^{-1}$}. In order to estimate uncertainties on the DEM, we computed Monte Carlo (MC) realizations of the data. We considered a 20\% uncertainty in the intensities, which takes account of uncertainties in the atomic data \citepads{2013A&A...555A..47D} and obtained 400 MC solutions by randomly varying the input intensities. We applied similar methodology given by  \citetads{2012ApJ...761...62C} to plot the MC solutions on the best-fit DEM solution. The blue rectangles represents 50\%, red rectangles 80\% and yellow rectangles 95\% of the MC solutions in each temperature bin (See fig.~\ref{fig9}). 

The EIS DEM curves are not very well constrained above \mbox{log \textit{T} [K] = 6.4}, since the highest temperature line observed was \mbox{Fe \textrm{XV} ($\lambda$284.160)}. No emission was observed from \mbox{Fe \textrm{XVII} ($\lambda$254.87)} at \mbox{log \textit{T} [K] = 6.75}. In the high temperature range (\mbox{$>$log \textit{T} = 6.5}), the DEM falls sharply. In the lower temperature range, there are no lines observed between \mbox{log \textit{T} [K] = 5.8 and log \textit{T} [K] = 6.1} in the EIS observing sequence used, so the DEM values in this range are less reliable.

\subsubsection{AIA-XRT DEM}    \label{section3.6.2}

\begin{figure}[!hbtp]
\begin{center}
\includegraphics[trim=3.0cm 12.0cm 0.1cm 4.5cm,width=0.52\textwidth]{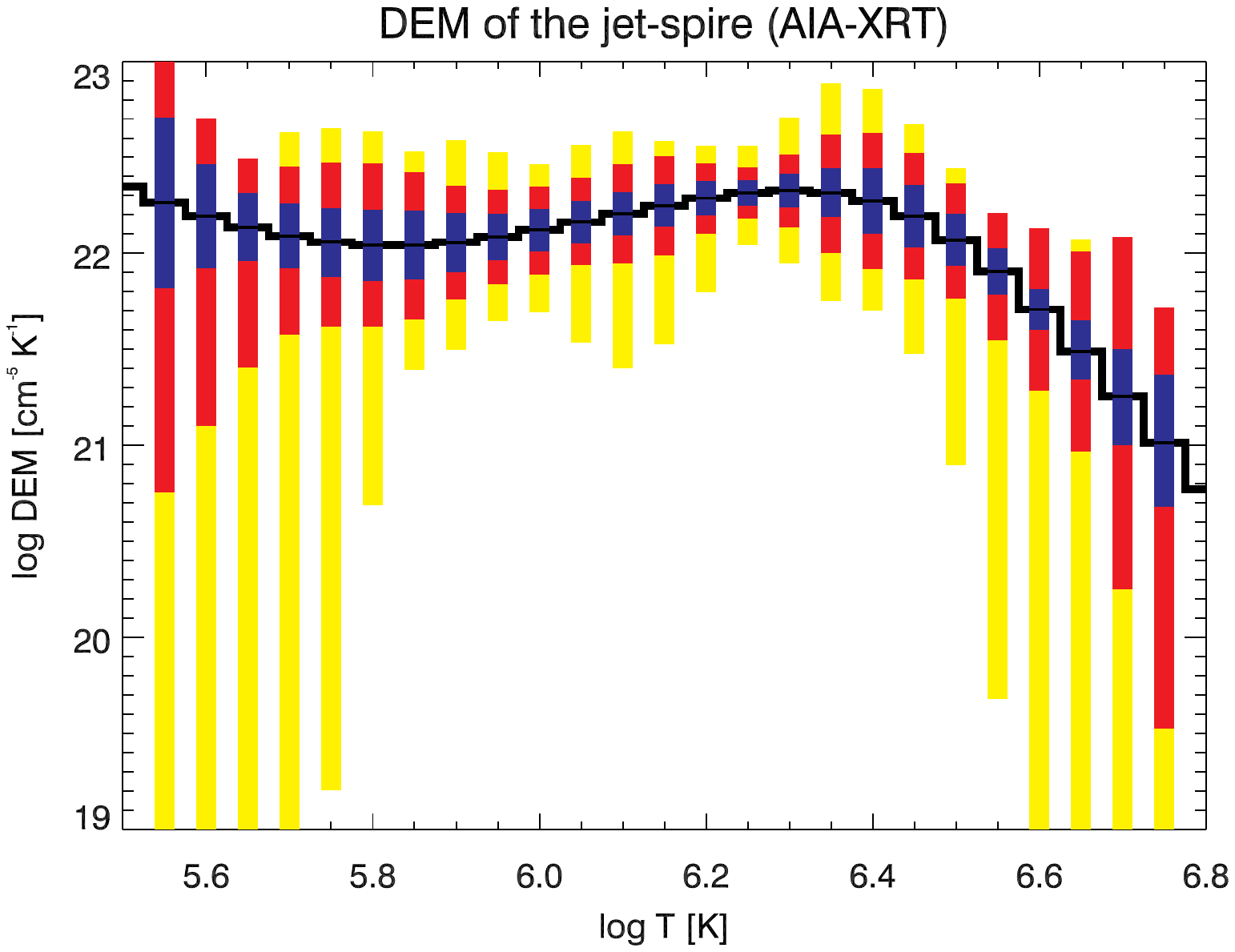} 
\includegraphics[trim=3.0cm 12.0cm 0.1cm 4.5cm,width=0.52\textwidth]{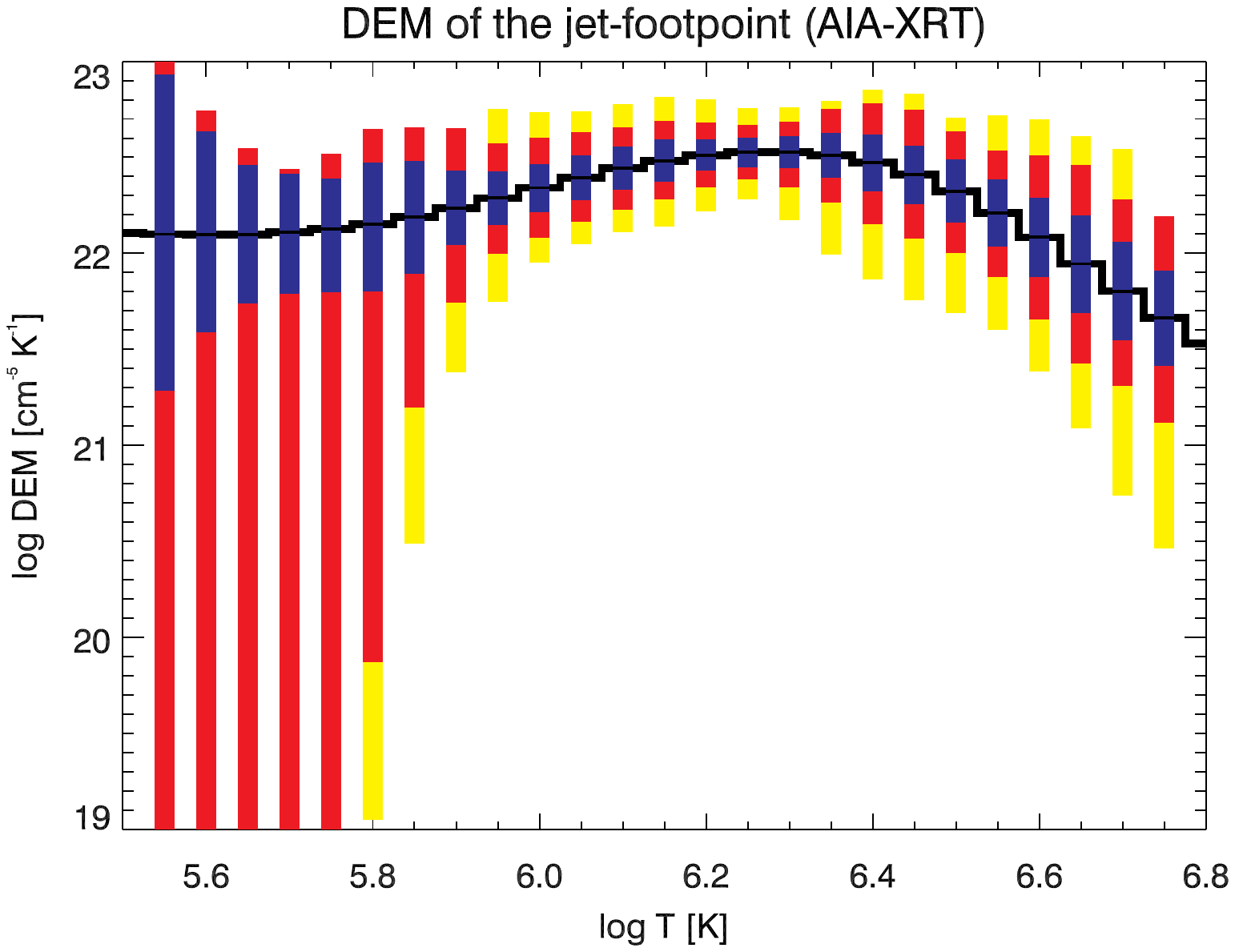} 
\caption{AIA-XRT DEM curves for the jet-spire (top panel) and jet-footpoint (bottom panel). The DEMs are obtained from the ‘XRT\_DEM\_ITERATIVE2’ method (black curve). The blue rectangles represent 50\%, red rectangles 80\% and yellow rectangles 95\% of the MC solutions in each temperature bin. \label{fig10}}
\end{center}
\end{figure}

Using the co-aligned XRT images from the Ti-poly filter together with time-averaged AIA images in six wavelength channels (94, 131, 171, 193, 211 and 335~{\AA}) which are sensitive to a range of coronal temperatures, we performed the DEM analysis again using the same method described in section~\ref{section3.6.1}. The purpose of using X-ray Ti-poly images in the DEM analysis is to constrain high temperature emission.

We extracted the same small region of 2$\arcsec$ $\times$ 3$\arcsec$ (2$\times$3 pixels) area around the jet spire and the footpoint which we used to obtained DEM curves from the EIS. The DN per sec per EIS pixel in each of the six time-averaged AIA channels and the \mbox{Ti-poly} filter were spatially averaged over all pixels in the region. We used these count rates as an input to compute DEM. For the inversion, we used a range of temperatures \mbox{log \textit{T} [K] = 5.5} to \mbox{log \textit{T} [K] = 7.1} with $\bigtriangleup$ log \textit{T} = 0.05 intervals.

The ‘xrt\_make\_wave\_resp.pro’ routine was used to produce the spectral responses for the Ti-poly filter. We obtained XRT and AIA temperature responses from the CHIANTI v.8 (\citeads{2015A&A...582A..56D}) atomic database using the ‘ISOTHERMAL’ procedure (see Appendix in \citeads{2011A&A...535A..46D}). In order to compare the DEM results from the EIS with those obtained by the AIA-XRT instruments, we calculated the responses ‘per EIS pixel’ (see fig~\ref{figa1} and fig~\ref{figa2} in the Appendix). We took into account the time-dependent degradation factor for these instruments as in the SSW. We used a set of ‘photospheric’ abundances (\citeads{2009ARA&A..47..481A}) along with the electron density obtained from the \mbox{Fe \textrm{XII}} line intensity ratio (see section \ref{section3.4}) in this calculation. 
 
Figure \ref{fig10} shows the DEM curves (black curves) for the jet-spire (top panel) and the jet-footpoint (bottom panel) respectively. The errors on the DEM curves were estimated by considering a 20\% uncertainty in the intensities, coming primarily from the atomic data. The blue rectangles represents 50\%, red rectangles 80\% and yellow rectangles 95\% of the MC solutions in each temperature bin. The DEM for the jet-spire peaks at \mbox{log \textit{T} [K] = 6.3} with \mbox{DEM = 2.1$\times$10$^{22}$ cm$^{-5}$ K$^{-1}$}, and for the jet-footpoint, it also peaks at \mbox{log \textit{T} [K] = 6.3} with \mbox{DEM = 3.4$\times$10$^{22}$ cm$^{-5}$ K$^{-1}$}. The error bars below \mbox{log \textit{T} [K] = 5.9} and above \mbox{log \textit{T} [K] = 6.4} are large, so the DEM values in this range are less reliable. \mbox{Table~\ref{table4}} shows the observed (column 2 and 5) and predicted (column 3, 4 and 6, 7) intensities (average DN/s per EIS pixel) in the region of the jet-spire and the jet-footpoint respectively. The intensities reproduced by the AIA-XRT DEM are closer to the observed intensities than the intensities reproduced by the EIS DEM. In the case of the AIA 94~{\AA} channel, the EIS DEM shows large differences in the observed and predicted intensities. Since the EIS observation only covers temperature between \mbox{log \textit{T} [K] = 5.7 and 6.4}, it is possible that the EIS DEM is unable to reproduce the high temperature emission that is observed by the AIA and XRT. We note that although the peak DEM values and temperature of peak DEM are similar for both the EIS and AIA-XRT DEMs, the high temperature values (\mbox{log \textit{T} [K] > 6.5}) for AIA-XRT do not fall as sharply as for EIS, although the scatter of MCMC results is large.



 \begin{table*}[!btp]
 \renewcommand\thetable{4} 
 \centering
\caption{Observed and predicted SDO/AIA count rates for the jet-spire and jet-footpoint obtained using photospheric abundances.}
 \resizebox{15cm}{!} {
 \begin{tabular}{lccccccc}

 \hline
 \hline
 &			&		 &			&		 &			&		 &\\ 	 	

 &{Band}		&{Observed} 	 &{Predicted}		&{Predicted}	&{Observed} 	 &{Predicted}		 &{Predicted}	\\  

 &({\AA})		&		 &{(AIA-XRT DEM)}	&{(EIS DEM)}	&		 &{(AIA-XRT DEM)}	 &{(EIS DEM)}			\\ 	 
  
 &			&{spire}	 &{spire}		&{spire}	&{footpoint}	 &{footpoint}		 &{footpoint}	\\  		

 \hline		
 &			& 		 &		 	&		&		 & 			 &\\ 	 	
				  
 &{AIA 94}		&45	 	 &27 (-40\%)		&10 (-78\%)	&75 	 	&67 (-11\%)		 &22 (-71\%)	\\	 	 	
 
 &{AIA 131}		&303	 	 &244 (-20\%)		&112 (-63\%)	&455 	 	&350 (-23\%)		 &366 (-20\%)	\\   
 
 &{AIA 171} 		&4747		 &5930 (25\%)		&2134 (-55\%)	&7651 	 	&7565 (-1\%) 		 &5867 (-23\%)	\\	  	 	

 &{AIA 193}		&4494		 &5144 (15\%)		&3181 (-29\%)	&8063 	 	&7558 (-6\%)	 	 &6942 (-14\%)	\\   	 	 	
 
 &{AIA 211} 		&1946	 	 &1984 (2\%)		&1461 (-25\%)	&2957 	 	&2976 (1\%)		 &2891 (-2\%)	\\	 	 	

 &{AIA 335}		&117		 &101 (14\%)     	&59 (-50\%)	&179 	 	&170 (-5\%)		 &116 (-35\%)	\\ 
 
 &			&		 &			&		 &			&		 &\\

 \hline
 \hline

  \end{tabular} 
}
\tablefoot{Column 2 and 5 indicate the observed count rates (averaged DN/s per EIS pixel) for the region of the jet-spire and the jet-footpoint respectively. Column 3, 4 and 6, 7 indicate predicted count rates obtained from the AIA-XRT DEM and EIS DEM for the region of the jet-spire and the jet-footpoint respectively. Values in parenthesis in columns 3,4 and 6,7 show the percentage differences between observed and predicted intensities.} \label{table4}

\end{table*}

\subsection{Emission measure and filling factor measurement} \label{section3.7}
 
The emission measure can be measured from individual spectral emission lines at their temperature of formation \citepads{1963ApJ...137..945P}. Considering a spectral line formed over a range of temperatures -  \mbox{log \textit{T}$_{max}$ - 0.15} to \mbox{log \textit{T}$_{max}$ + 0.15} (where \textit{T}$_{max}$ is the temperature where the contribution function has its maximum), we can determine the emission measure by integrating the DEM values within that temperature range. 

In this analysis, we consider the Fe \textrm{XII} line formed over a temperature range from \mbox{log \textit{T} [K] - 6.05 to 6.35} and we calculated emission measure from the EIS DEM and also from the \mbox{AIA-XRT} DEM in the region of the spire and the footpoint.  

The column depth for the spire and the footpoint is estimated from the size of the observed structure in the EIS raster images. In this study, we observed the emission from the spire and the footpoint over two consecutive EIS raster positions i.e. for 2\arcsec. By assuming that the structure has a cylindrical geometry, we calculated a filling factor using equation~\ref{eq:phi}. The filling factor was 0.02 in both regions using photospheric abundances and 0.005 for coronal abundances.

Table~\ref{table5} displays the results for plasma parameters. The values obtained from \mbox{AIA-XRT} DEM are given in column 2 and 4 and the values obtained from the EIS DEM curves are given in column 3 and 5 respectively. 

\subsection{Synthetic spectrum} \label{section3.8}

Synthetic spectra were calculated from the EIS DEM and AIA-XRT DEM curves in the region of the spire and the footpoint. The purpose was to see whether the DEM curves predict any high-temperature emission, such as emission from \mbox{Fe \textrm{XVII} ($\lambda$254.87)} at \mbox{log \textit{T} [K] = 6.75} in the region of the spire or the footpoint (even though EIS did not show any evidence of this). 

We used the ‘CH\_SS’ CHIANTI routine to obtain the spectra. We calculated synthetic spectra in the SW and LW channels of the EIS instrument using the same set of parameters (abundances and densities; see table \ref{table5}). We adopted the radiometric calibration by \citetads{2013A&A...555A..47D} to calculate the EIS effective areas. The synthetic spectra were then convolved with the effective areas of both channels and the contribution of spectral lines and continuum emission was determined (see the Appendix for details). 

The synthetic spectra obtained from the EIS DEM and AIA-XRT DEM curves predict similar lines. Small differences were seen in the intensities which could be explained by the highly variable nature of the jet structure. The AIA lightcurves in these regions (see figs. \ref{fig4} and \ref{fig5}) show changes in the count rates in all AIA channels during the EIS raster observations. Also, we have to take into account the spatial resolution and exposure times for EIS, AIA and XRT instruments, which also has an effect on the DEM curves.

The SW and LW channels show more count rates in the case of the jet footpoint compared to the spire structure. The \mbox{Fe \textrm{XII}} ($\lambda$195.119) line dominates the SW channel; whereas \mbox{Fe \textrm{XIV} ($\lambda$264.7889)} and \mbox{Fe \textrm{XV}} ($\lambda$284.160) dominate the LW channel in the synthetic spectra. Neither synthetic spectra for the spire predict any measurable contribution from the high temperature lines such as \mbox{Fe \textrm{XVII}} \mbox{($\lambda$254.87)} at \mbox{log \textit{T} [K] = 6.75} confirming that the high temperature part of the DEM is consistent with the observed EIS spectra. In the case of the footpoint region, both synthetic spectra predict weak contributions from \mbox{Ca \textrm{XVII} ($\lambda$192.85)} (\mbox{log T [K] = 6.7}) and \mbox{Fe \textrm{XVII} \mbox{($\lambda$254.87)}} lines. The predicted \mbox{Fe \textrm{XVII}} \mbox{($\lambda$254.87)} signal is weak but inconsistent with the EIS observation, which shows no signal.


%




 \begin{table*}[!btp]
 \renewcommand\thetable{5} 
 \centering
 \caption{Plasma parameters calculated using EIS data and DEM curves}
 \resizebox{12cm}{!} {
 \begin{tabular}{lccccc}

 \hline
 \hline
&&&&&\\

 &{Event}	        			&\multicolumn{2}{c}{Photospheric Abundances} 	 	&\multicolumn{2}{c}{Coronal Abundances}		\\ 
 
 &&&&&\\
 
 &						&AIA-XRT DEM 		&EIS DEM			&AIA-XRT DEM 	 	&EIS DEM		\\ 

 &&&&&\\
 
 \hline
 
 &&&&&\\
 
 &{jet-spire} 					&   			&		 	 	 & 			&  		\\ 
 
 &{log T (K) for peak DEM}			&6.3			&6.3			 	 &6.3			&6.3	  	\\

 &{peak DEM (cm$^{-5}$ K$^{-1}$)} 		&2.1$\times$10$^{22}$ 	&2.1$\times$10$^{22}$	 	 &5.3$\times$10$^{21}$ 	&5.5$\times$10$^{21}$	\\ 

 &{EM for Fe XII (cm$^{-5}$ )}			&2.2$\times$10$^{28}$ 	&1.7$\times$10$^{28}$	 	 &5.4$\times$10$^{27}$	&4.3$\times$10$^{27}$	\\
 
 &{width of the spire (cm)}			&			&1.5 $\times$10$^{8}$ 	 	 &			&	  	\\
 
 &{density from Fe \textrm{XII} (cm$^{-3}$)}	&			&7.6$\times$10$^{10}$		 &			&	  	\\

 &{filling factor from EIS}			&			&0.02				 &			&0.005		\\
 
 &						&			&	 			 &			&	  	\\
 
 \hline
 \hline
 
 &						&			&	 			 &			&	  	\\
 
 &{jet-footpoint}  				&			&	 			 &			&		\\
 
 &{log T (K) for peak DEM}			&6.3			&6.3			 	 &6.3			&6.3		\\

 &{peak DEM (cm$^{-5}$ K$^{-1}$)}		&3.4$\times$10$^{22}$ 	&3.9$\times$10$^{22}$ 		 &9.5$\times$10$^{21}$ 	&1.0$\times$10$^{22}$ 	\\ 
 
 &{EM for Fe XII (cm$^{-5}$)}			&3.5$\times$10$^{28}$ 	&3.4$\times$10$^{28}$	 	 &9.6$\times$10$^{27}$ 	&8.6$\times$10$^{27}$	\\
  
 &{width of the footpoint (cm)}			&			&1.5 $\times$10$^{8}$		 &			&	  	\\
 
 &{density from Fe \textrm{XII} (cm$^{-3}$)}	&			&1.1$\times$10$^{11}$		 &			&	  	\\
 
 &{filling factor from EIS}			&			&0.02				 &			&0.005		\\
 
 &						&			&				 &			&	  	\\
 
 \hline
 \hline

  \end{tabular} 
} 
\label{table5}

\end{table*}

\begin{figure*}[!btp]
\begin{center}
\includegraphics[width=0.9\textwidth]{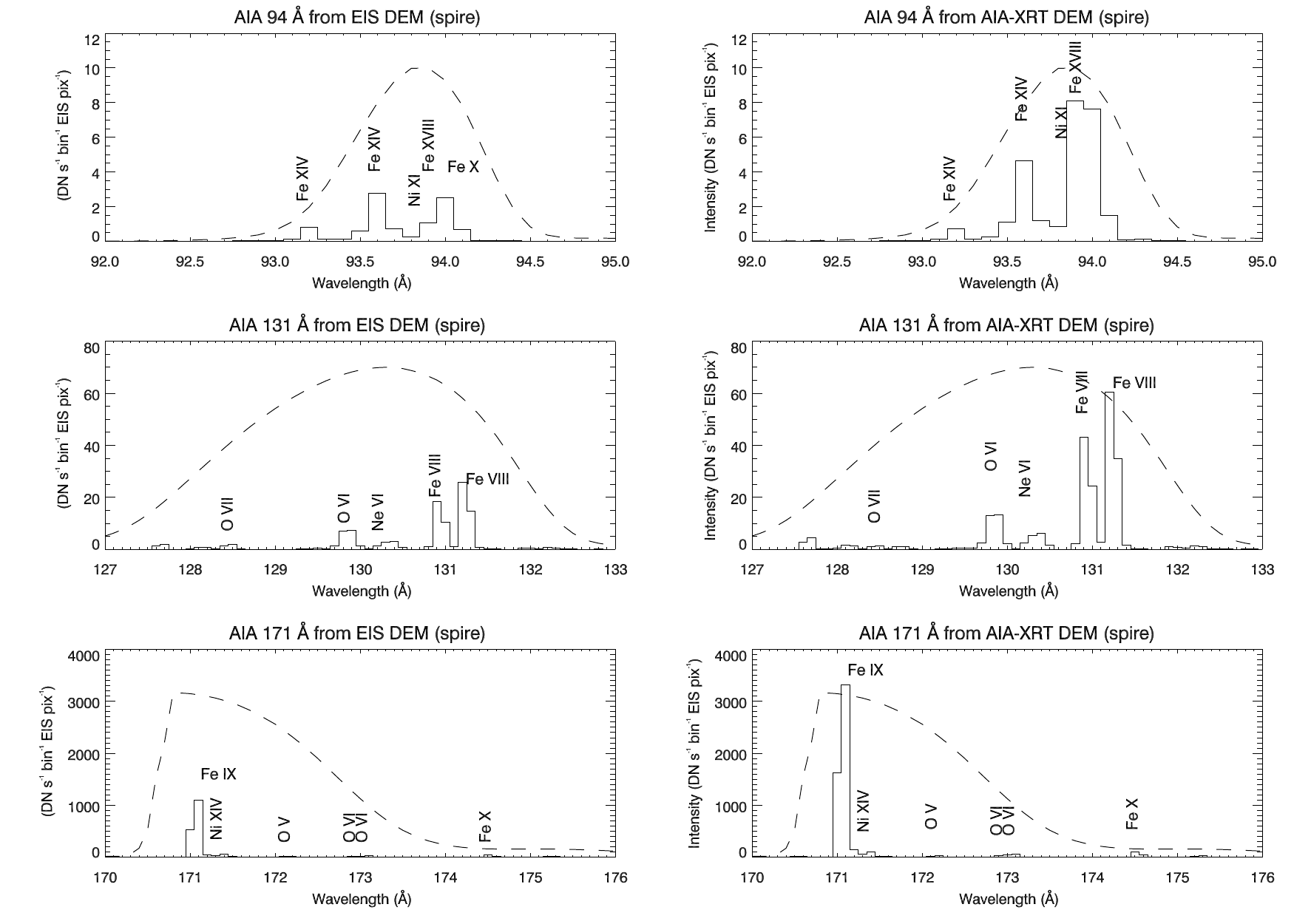}
\includegraphics[width=0.9\textwidth]{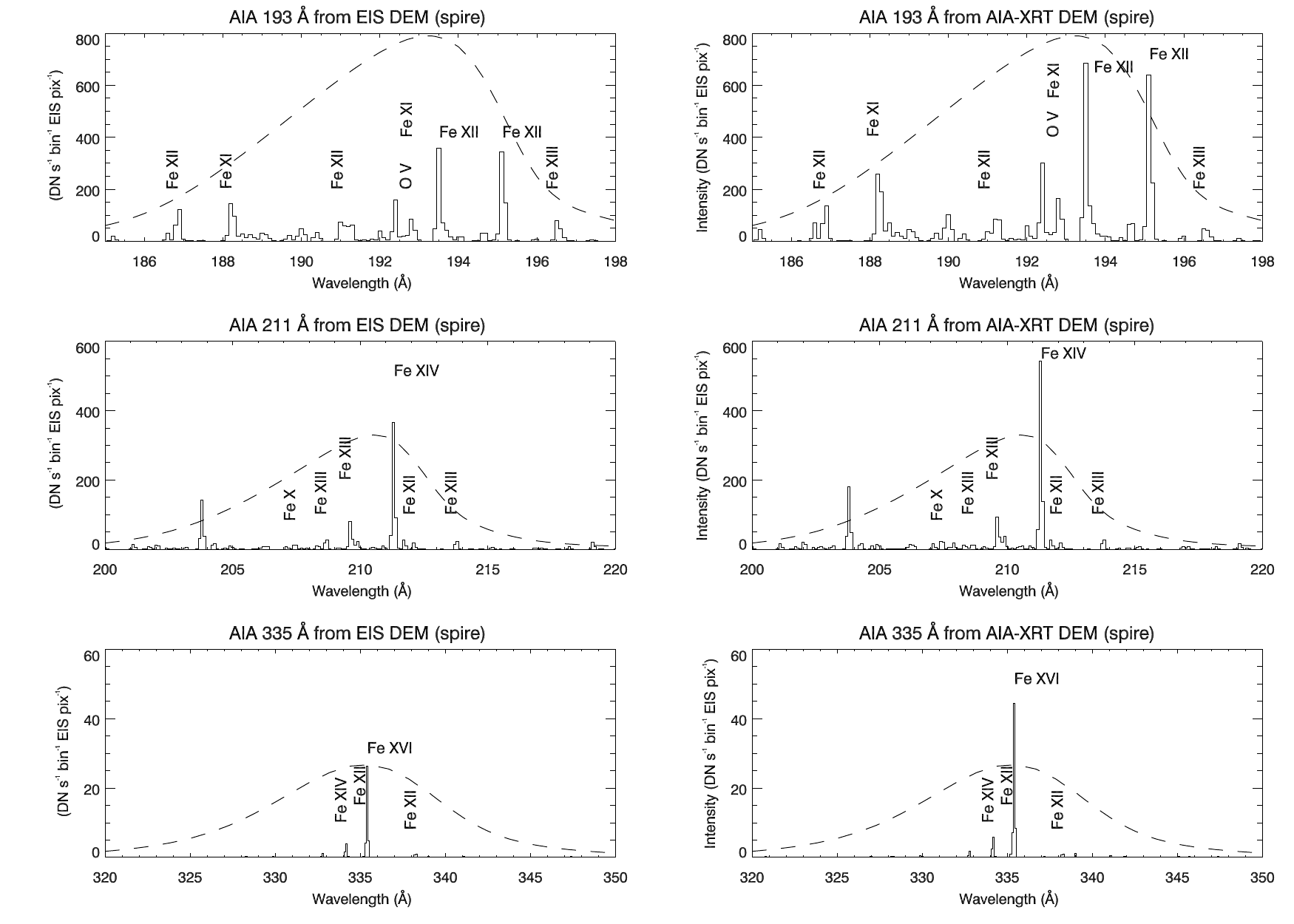}
\caption{The synthetic spectra for each AIA channel obtained from the EIS DEM (left panel) and the AIA-XRT DEM (right panel) for the region of the spire. The overplotted dashed lines show the effective area (scaled) of each AIA channels. \label{fig11}}
\end{center}
\end{figure*}


\begin{figure*}[!btp]
\begin{center}
\includegraphics[width=0.9\textwidth]{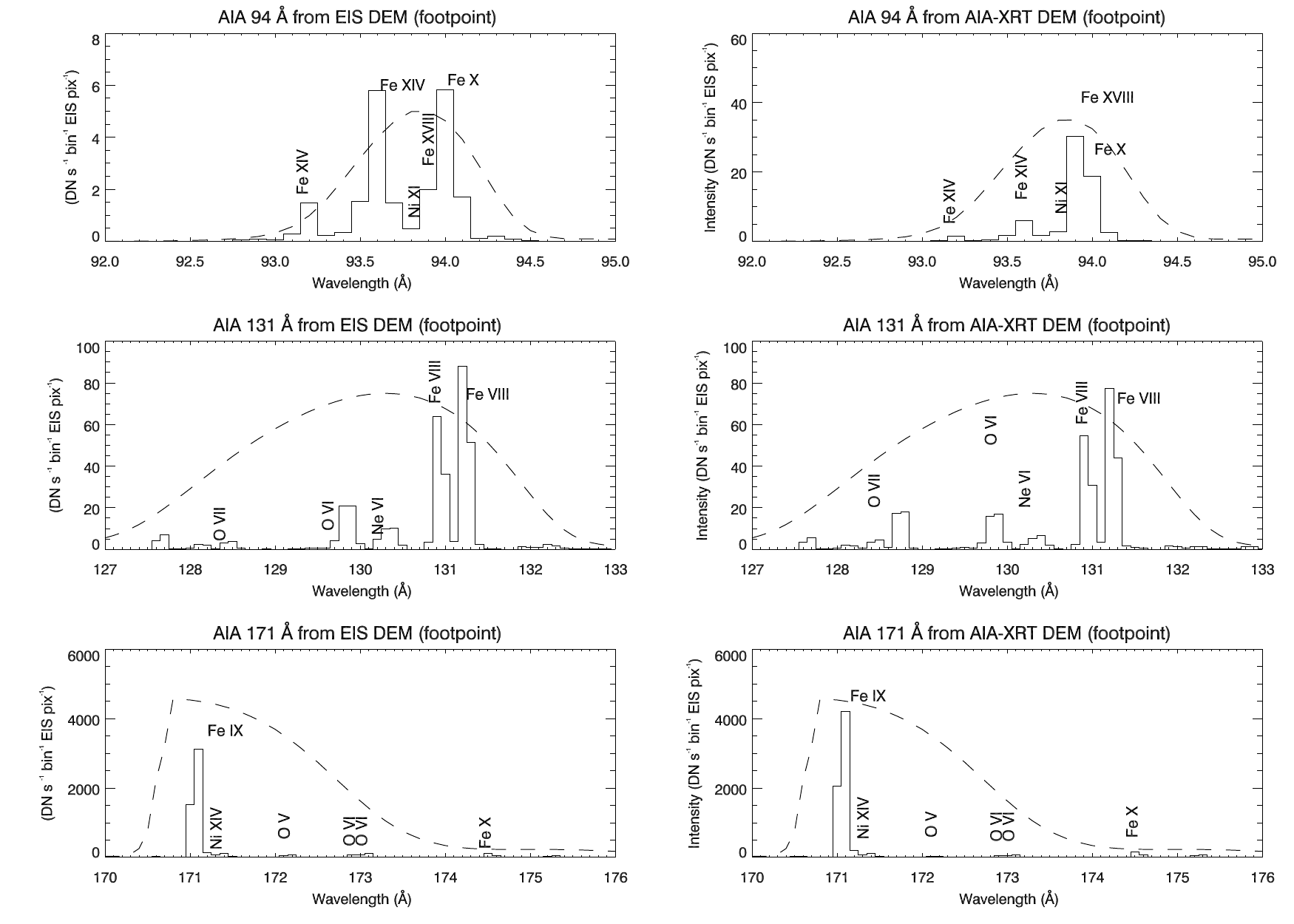}
\includegraphics[width=0.9\textwidth]{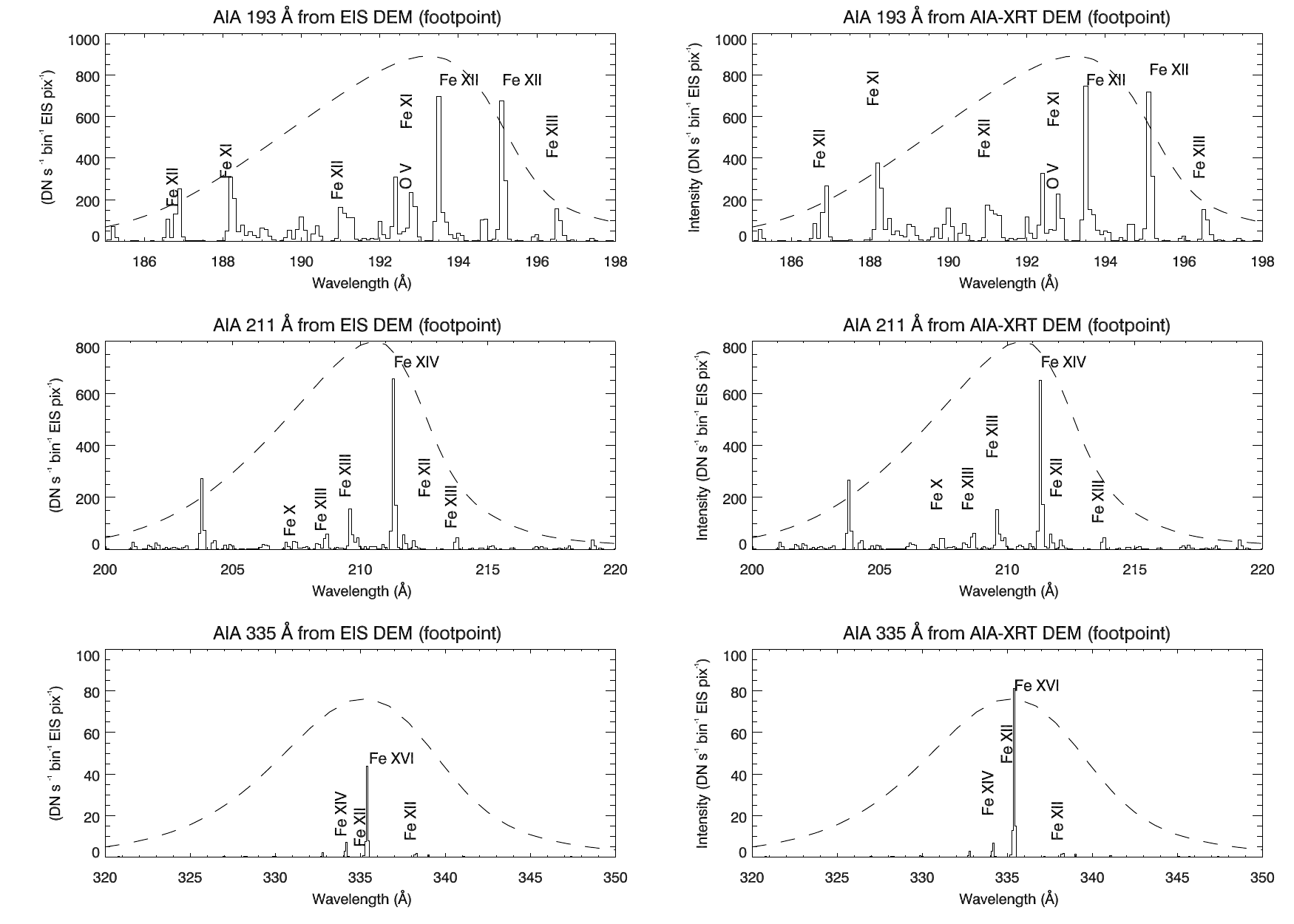}
\caption{The synthetic spectra for each AIA channel obtained from the EIS DEM (left panel) and the AIA-XRT DEM (right panel) for the region of the footpoint. The overplotted dashed lines show the effective area (scaled) of each AIA channels. (The Y-axis scale is different in the case of the AIA 94~{\AA} channel for the intensities predicted from the EIS DEM and the AIA-XRT DEM.)\label{fig12}}
\end{center}
\end{figure*}

\subsection{Effect of varying the temperature range and abundances in the DEM analyses} \label{section3.9}

The effects of varying the temperature range and elemental abundances in the DEM analyses was investigated. We followed the same procedure given in section~\ref{section3.6} to obtained DEM from AIA-XRT and the EIS observations using coronal abundances by \citetads{1992PhyS...46..202F} and a temperature range from \mbox{log \textit{T} [K] = 5.5 to 7.1}. We obtained similar curves (in shape) as shown in fig. \ref{fig9} and \ref{fig10}. The jet-spire and the jet-footpoint had showed the same peak temperature (see table~\ref{table5}). However, the peak DEM, EM and filling factor values obtained using the coronal abundances were found to be a factor of four lower than the values obtained using the photospheric abundances.

\subsection{AIA count rates}

The EIS DEM curves were used to predict the total count rates in each AIA channel. We used the ‘CH\_SS’ CHIANTI routine to obtain synthetic spectra in each of the AIA channel. The effective areas of each AIA channel were calculated ‘per EIS pixel’. We examined the contribution of spectral lines and continuum emission to each of the channel. The results confirm the multi-thermal emission contributing to the AIA channels in both regions, the only exception being the AIA 171~{\AA} channel.

Figure \ref{fig11} and \ref{fig12} shows the predicted total count rates in each AIA channel obtained from the EIS DEM curves for the region of the spire and the region of the footpoint respectively. The differences seen in the observed and predicted count rates can be attributed to the variability seen in the jet structure (see figs. \ref{fig4} and \ref{fig5}). Count rates in each AIA channel are found to be higher in the region of the footpoint than in the region of the spire (see Table \ref{table4}).

\subsection{The evolution of the footpoint}

\begin{figure*}[!btp]
\begin{center}
\includegraphics[trim=1.5cm 12.5cm 1.2cm 4.0cm, width=1.0\textwidth]{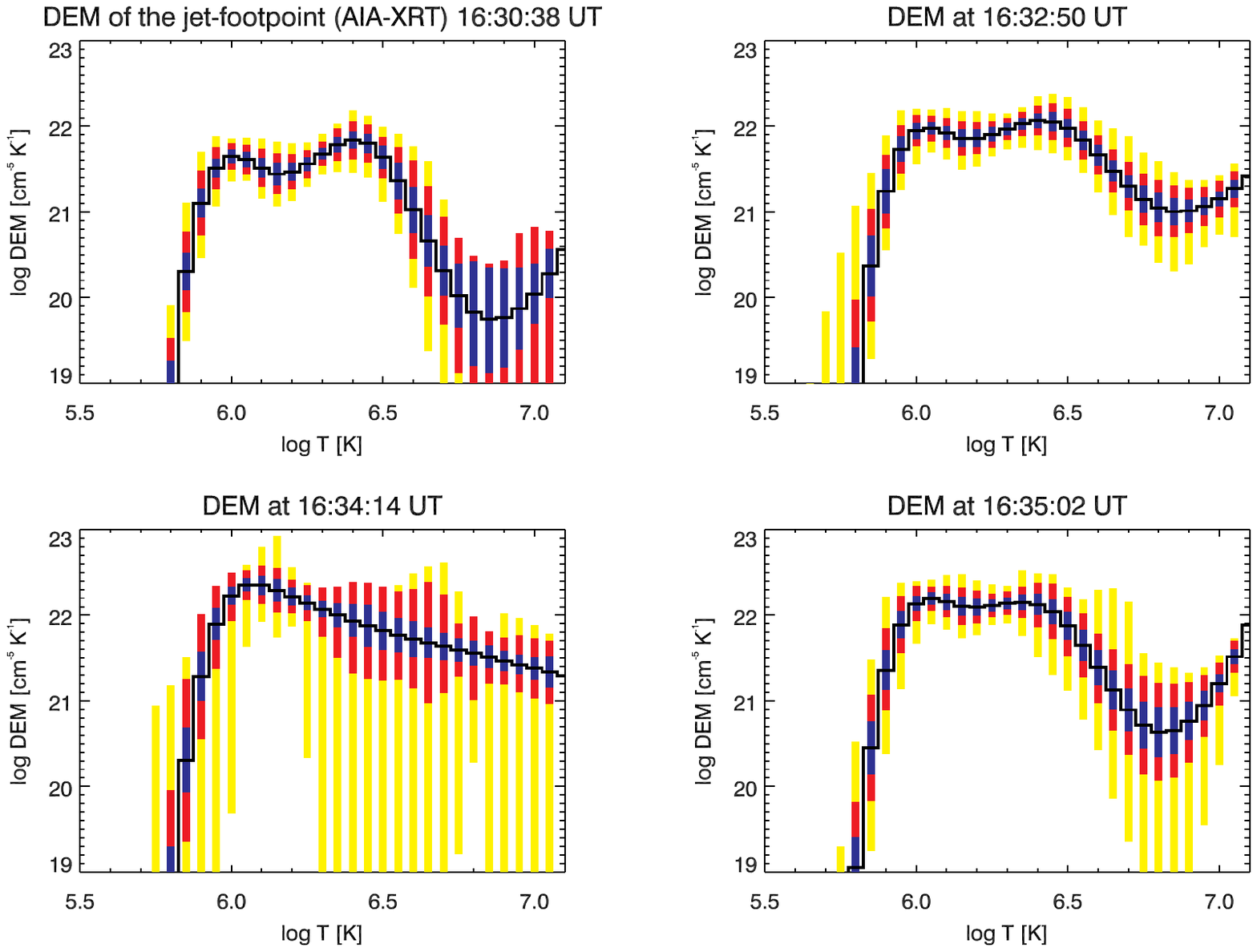} 
\caption{The DEM results obtained from the AIA and XRT images during the temporal evolution of the footpoint region. \label{fig13}}
\end{center}
\end{figure*}


\begin{figure*}[!btp]
\begin{center}
\includegraphics[trim=1.5cm 0.5cm 2.8cm 8.5cm, width=1.1\textwidth]{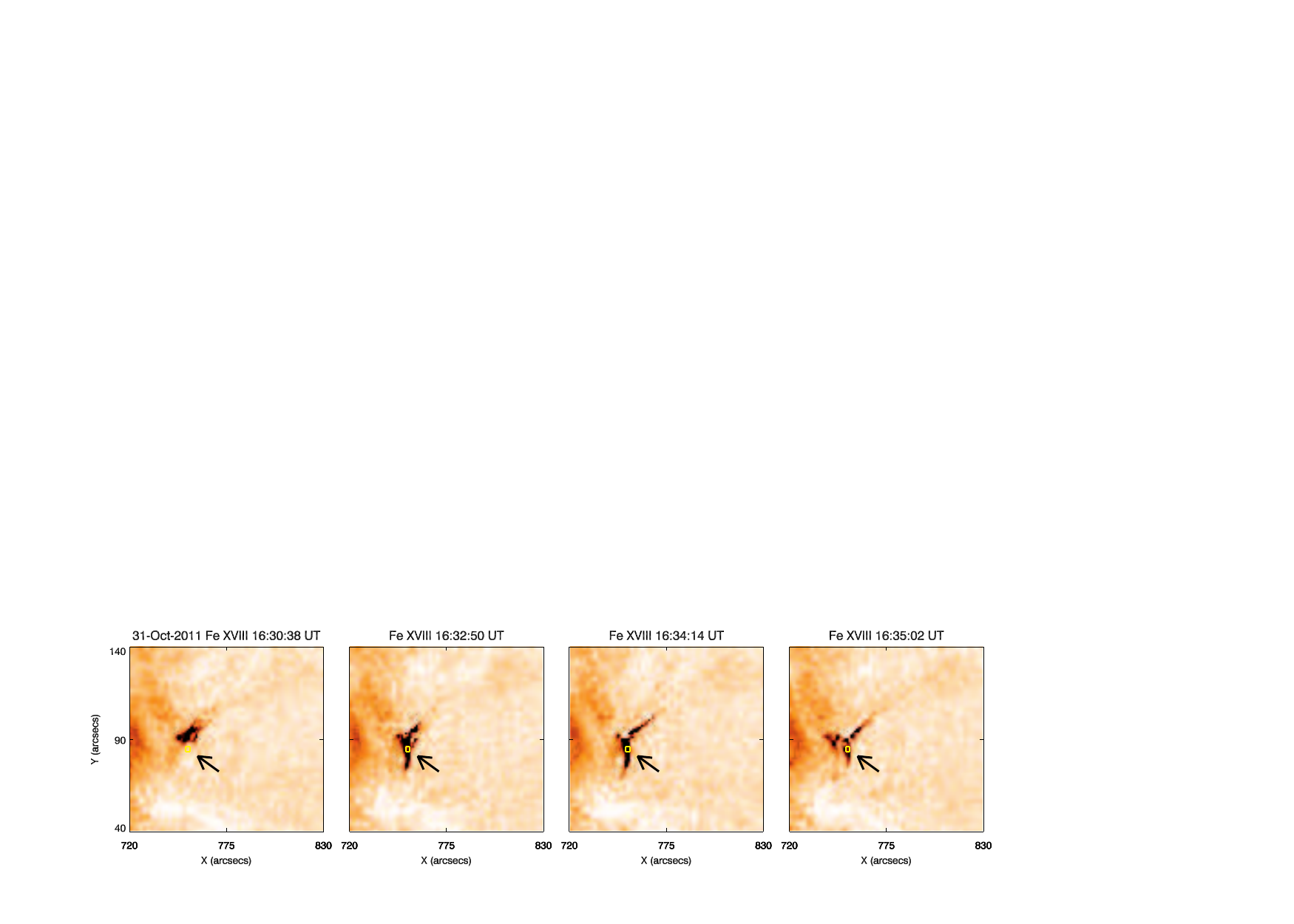}
\caption{\mbox{Fe \textrm{XVIII} (93.932~{\AA})} maps during the evolution of the footpoint region. The yellow over plotted box (shown by black arrows) shows the same footpoint region as shown in fig. \ref{fig5} \label{fig14}}
\end{center}
\end{figure*}


The continuous coverage of the jet with SDO/AIA and Hinode/XRT enabled us to investigate further the temporal evolution of the footpoint region. \mbox{Figure \ref{fig5}} (bottom panel) shows the temporal evolution of the footpoint region in all AIA coronal channels and its variable nature is clearly seen over the period of time. Before the EIS raster times (\mbox{16:35:08} and \mbox{16:35:30} UT - shown by black lines in fig. \ref{fig5} bottom panel), the light curves showed a peak in the AIA 94~{\AA} and 335~{\AA} channels at 16:34 UT. These peaks were followed by peaks in the 211~{\AA}, the 193~{\AA}, and finally the 171 and 131~{\AA} emission. This scenario indicates that the hot temperature cooled down over the period of time from \mbox{log \textit{T} [K] = 6.6} and that EIS observed the jet-footpoint during its cooling phase.

In order to investigate this further, we performed a DEM analysis using the AIA and XRT images. We have taken the AIA images (0.6\arcsec per pixel) which are closest to the XRT timings and degraded them to the XRT resolution (1.02$\arcsec$). We performed the DEM analysis for every 1 min of observations from \mbox{16:30 UT} to \mbox{16:40 UT} in the same box at the region of the footpoint (shown in the fig. \ref{fig5} top panel). \mbox{Figure \ref{fig13}} shows the DEM curves at four different times during the evolution of the footpoint. We observed significant differences in the DEMs for the temperature range from \mbox{6.5 < log \textit{T} [K] < 6.9}. At \mbox{16:30 UT} (top left panel), the DEM peaks at \mbox{log \textit{T} [K] = 6.4} and falls sharply until \mbox{log \textit{T} [K] = 6.9}. At \mbox{16:32 UT} (top right panel), the DEM also peaks at \mbox{log \textit{T} [K] = 6.4} but the DEM values are increased for the temperatures \mbox{> log \textit{T} [K] = 6.6}. The shape of the DEM changes drastically at \mbox{16:34 UT} (bottom left panel), where it peaks at \mbox{log \textit{T} [K] = 6.1} and gradually decreases until \mbox{log \textit{T} [K] = 7.0}. The DEM values at a temperature around \mbox{log \textit{T} [K] = 6.7} has increased rapidly over the period of two minutes. With time, the DEM decreases again for the temperature \mbox{> log \textit{T} [K] = 6.5} at \mbox{16:35 UT} (bottom right panel). These DEM results indicate that there is a rise in temperature then a gradual cooling phase following the peak at \mbox{16:34 UT}. This scenario is consistent with the nature of the observed AIA light curves.

\subsection{Estimation of Fe XVIII emission in the jet-footpoint}

During the evolution of the jet-footpoint, the DEM curves showed the presence of temperature \mbox{> log \textit{T} [K] = 6.7} (see \mbox{fig. \ref{fig13}}). In order to investigate the validity of this high temperature, we checked whether there is a significant presence of \mbox{Fe \textrm{XVIII}} emission in the footpoint region. For the AIA 94~{\AA} channel, we obtained \mbox{Fe \textrm{XVIII}} (93.932~{\AA}) maps using the empirical combination of the AIA 94, 211 and 171~{\AA} channels given by \citetads{2013A&A...558A..73D}. \mbox{Figure \ref{fig14}} shows the \mbox{Fe \textrm{XVIII}} images at the same timings as those for which we have obtained DEMs (see fig. \ref{fig13}) during the evolution of the footpoint. These images clearly show the \mbox{Fe \textrm{XVIII}} emission in the footpoint region at \mbox{16:32}, \mbox{16:34} and \mbox{16:35 UT}. The emission from \mbox{Fe \textrm{XVIII}} was not seen at the same small box region at \mbox{16:30 UT}, before the rise in temperature of the footpoint.

We estimated \mbox{Fe \textrm{XVIII}} averaged count rates in the AIA 94~{\AA} channel in the same region (shown by yellow box in fig. \ref{fig14}) and compared them with predicted averaged count rates obtained from the DEM (see fig. \ref{fig13}). Table~\ref{table6} shows a very good agreement between the \mbox{Fe \textrm{XVIII}} estimated and predicted count rates at \mbox{16:32}, \mbox{16:34} and \mbox{16:35 UT}.


 \begin{table}[!hbtp]
 \renewcommand\thetable{6} 
 \centering
\caption{Estimated and predicted \mbox{Fe \textrm{XVIII}} count rates (averaged DN/s) for the AIA 94~{\AA} channel obtained from the AIA-XRT DEM (see fig. \ref{fig14}) using the photospheric abundances for the jet-footpoint region shown in \mbox{fig. \ref{fig13}}.}
 \resizebox{7cm}{!} {
 \begin{tabular}{lccc}

 \hline
 \hline
 &		&		 	&		  		\\ 

 &{Time}	&{Estimated} 	 	&{Predicted}		\\  
 
 &{(UT)}	&{(from AIA 94~{\AA})}	&{(from DEM)}				\\  		

 \hline		
 &		& 		 	&		 		\\ 	 	
			  
 &{16:30:38}	&0.4	 	 	&3				\\	 	 	
 
 &{16:32:50}	&17	 	 	&20 				\\         	

 &{16:34:14} 	&34	 	 	&28				\\	  	 	

 &{16:35:02}	&12 	 	 	&18				\\   	 	 	
   
 &		& 		 	&		 		\\ 	 	

 \hline
 \hline

  \end{tabular} 
}
\label{table6}

\end{table}

\section{Discussion and Summary} \label{section4}

In this paper, we present a comprehensive investigation of the temperature structure of the jet-spire and the jet-footpoint of a recurrent AR jet observed on October 31, 2011 using simultaneous imaging (from the AIA and XRT) and spectroscopic observations (from the EIS instrument). The jets originated from the western edge of AR NOAA 11330 \mbox{(N08 W49)}. The highly variable nature of the jet-spire and the footpoint was observed during the jet evolution (see figs. \ref{fig4} and \ref{fig5}). We also observed plasma-blobs moving along the jet-spire in the AIA channels (see online movie1.mp4).

We studied the temperature structure of the jet-spire and the jet-footpoint by performing a DEM analysis (see section~\ref{section3.6}). The plasma along the line-of-sight in the jet-spire and jet-footpoint was found to be peak at 2.0 MK (\mbox{log \textit{T} [K] = 6.3}) and we obtained similar DEM values at the peak of DEM curves (see table~\ref{table5}). The EIS and \mbox{AIA-XRT} DEM curves in both regions appear to be in good agreement in the temperature interval from \mbox{log \textit{T} [K] = 5.9 - 6.3}. Substantial variations were found between solutions obtained from the MC iterations at lower temperatures. We note that the DEM curves are not well constrained below \mbox{log \textit{T} [K] = 5.8} and above \mbox{log \textit{T} [K] = 6.4}.

There are various factors which can affect the DEM results such as the choice of elemental abundances, range of temperatures over which the DEM inversion is performed, uncertainties in the atomic data and cross-calibration of instruments. We investigated cross-calibration issues by performing a similar analysis shown in section~\ref{section3.6} on a moss region (shown as yellow box in fig. \ref{fig1}) for which there was very little variation of the intensity with time (see Appendix). Based on the results obtained from the moss, we have confidence in the calibration of EIS, AIA and XRT. These results also confirm that the method we used in section \ref{section3.6} for the DEM analysis combining AIA and XRT observation is reliable.

The synthetic spectra for the spire do not predict any measurable contribution from the high temperature lines such as \mbox{Fe \textrm{XVII}} \mbox{($\lambda$254.87)} at \mbox{log \textit{T} [K] = 6.75} confirming that the high temperature part of the DEM is consistent with the observed EIS spectra. In the case of the jet-footpoint, both synthetic spectra predict weak contributions from \mbox{Ca \textrm{XVII}} ($\lambda$192.85) and  \mbox{Fe \textrm{XVII}} \mbox{($\lambda$254.87)}. With further investigation, we confirmed that there was emission from the \mbox{Fe \textrm{XVIII} (93.932~{\AA})} lines in the region of the footpoint during the early stages of the jet. We also found a good agreement between the estimated and predicted \mbox{Fe \textrm{XVIII}} count rates. The consistency between the nature of the AIA light curves and changes in the DEM values with temperatures during the evolution of the footpoint and the emission from the \mbox{Fe \textrm{XVIII} (93.932~{\AA})} line leads us to conclude that the hot component in the footpoint region was present initially that the jet had cooled down by the time EIS observed it. 

It is important to note that the predicted lines in each AIA channels confirmed the multi-thermal emission contributing to the AIA channels in the region of the spire and the footpoint (see fig. \ref{fig11} and \ref{fig12}). This is the first such detailed investigation of this nature of AR jets.

We calculated an electron density using \mbox{Fe \textrm{XII}} ($\lambda$186/$\lambda$195) line ratio density diagnostics. The electron density was found to be \mbox{\textit{N}$_\textrm{e}$ = 7.6$\times$10$^{10}$ cm$^{-3}$} at the region of the spire and \mbox{1.1$\times$10$^{11}$ cm$^{-3}$} at the region of the footpoint (see fig.\ref{fig7}) taking account of 20\% error, the density could be even higher. For the first time in AR jet studies, we observed a region (shown by white arrow in \mbox{fig. \ref{fig7} (b)}, bottom panel) which has \mbox{Fe \textrm{XII}} \mbox{$\lambda$186/$\lambda$195} ratio greater than 1.2. This indicates that the region has a high density (\mbox{log \textit{N}$_\textrm{e}$ $>$ 11.5}), which is above the range of sensitivity of \mbox{Fe \textrm{XII}} lines ratio diagnostics.

We also calculated a plasma filling factor using equation~\ref{eq:phi}. It was found to be 0.02 in the region of the spire and also in the region of the footpoint. These values are in agreement to the values obtained by \citeads{2008A&A...481L..57C} and \citeads{2011RAA....11.1229Y} from their spectroscopic study of a recurrent active region jet. The peak DEM, EM and filling factor values obtained using coronal abundances were found to be a factor of four lower than the values obtained using the photospheric abundances (see Table \ref{table5}). 

The \mbox{AIA 131~{\AA}} image showed a similar structure and morphology of the jet to the EIS \mbox{Fe \textrm{VIII}} ($\lambda$186.605; \mbox{log \textit{T} [K] = 5.8}) observations (see fig.~\ref{fig6}). We confirm the emission observed in the spire and the footpoint in the AIA 131~{\AA} images has a main contribution from the lower temperature \mbox{Fe \textrm{VIII}} line. We obtained an initial velocity of \mbox{524 km/s} for the jet observed at \mbox{14:58 UT} from the time-distance analysis (see section~\ref{section3.5}). 

A recent multiwavelength study of twenty AR jets by \citetads{2016arXiv160200151M} showed that most of the AR jets originated at the western periphery of the active region in the vicinity of sunspots and that they were temporally associated with nonthermal \mbox{type-III} radio bursts. The energetic particles gyrating along the open magnetic field structures generally produce nonthermal \mbox{type-III} bursts in the radio dynamic spectrum. Using a Potential Field Source Surface (PFSS) technique, authors investigated the spatial co-relation between the AR jets and \mbox{type-III} radio bursts and confirmed the presence of open magnetic field lines at the same region of the jet-footpoint. The velocities and DEMs found in the current analysis are consistent with their results.

There are very few simulations of jets which give actual quantitative values for the plasma properties that can be compared to observations. Most simulations tend to predict much higher temperatures and lower densities than we observed, however most of them related to coronal hole jets. Two studies addressed active region jets. 

\citetads{2009A&A...506L..45G} studied the interaction of an emerging bipole  and small active region by performing 3D MHD numerical simulations. They reported a hot ($\sim$2 MK) and high velocity (V = $\sim$100 km/s) bidirectional flows as a result of reconnection. They also observed a change in the shape and direction of the jet during the process. They discussed two scenarios of the jet event : firstly, they observed a ‘L-shaped’ reconnection jet moving with speeds of about \mbox{100 km/s} and reached a temperature around \mbox{1 MK}.  Later, the jet was found to be trapped in the ambient field and adopted an ‘arc-like’ shape. The jet started moving laterally and reached a temperature of about \mbox{2 MK}. The authors  found good qualitative and quntitative agreement between observations and simulations. 

\citeads{2010A&A...512L...2A} studied the long-term evolution of a similar system (a recurrent active region jet) by solving the time-dependent, resistive MHD equations in 3D. The authors reported recurrent jets which occurred in direction perpendicular to each other as a result of repeated reconnection events. The recurrent jets appeared to have different physical properties and it changed over the period of time. During the evolution, the authors reported an enhancement of the temperature along the reconnection
outflow and they also observed spikes along the leading edge of the jet which further moved along the parallel field lines. The successive reconnection events found to be less effective than the previous one and eventually the system attained an equilibrium. 

From our spectroscopic and imaging observations, we have seen an inverted-Y topology at the footpoint of the jet and an  arc-shaped jet-spire (see online movie1.mp4). Our observations show good agreement with the geometric shape of the jet and temperatures ($\sim$2 MK) seen in the numerical simulations. However, the bi-directional flow was not observed during the recurrent phase of the jet. The measured densities (spire = 10$^{10}$ and footpoint = 10$^{11}$) and velocities (\mbox{$\sim$524 km/s}) are found to be higher than the values reported in both simulations.  

It remains to be seen with new numerical simulations (which will be the subject of a future paper) if such dynamic behaviour can be explained within the standard reconnection scenario associated with flux emergence. 

It is important to note that AR jets are relatively common, and the fact that there is strong evidence that they occur in regions that are open to the heliosphere (\citeads{2011A&A...531L..13I}, \citeads{2015MNRAS.446.3741C}, \citeads{2016arXiv160200151M}) makes them one of the best candidates for future detailed studies with the Solar Orbiter and Solar Probe plus suite of remote-sensing and in-situ instruments. Further detailed studies of these events are therefore very useful to prepare for future observations with these missions.

\begin{acknowledgements}
The authors are grateful to the referee for the comments which helped to improve the manuscript significantly. SMM acknowledges support from the Cambridge Trust, University of Cambridge, UK. HEM and GDZ acknowledge the support of STFC. The authors also thank Mr. Paul Wright from University of Glasgow, UK for his valuable comments. AIA data are courtesy of SDO (NASA) and the AIA consortium. CHIANTI is a collaborative project involving George Mason University, the University of Michigan (USA) and the University of Cambridge (UK). Hinode is a Japanese mission developed and launched by ISAS/JAXA, with NAOJ as domestic partner and NASA and STFC (UK) as international partners. It is operated by these agencies in co-operation with ESA and NSC (Norway).

\end{acknowledgements}


\bibliographystyle{aa-note} 
  \bibliography{eis_jet_r2}    

  \newpage
\begin{appendix} \label{appendix}

\section{AIA responses \& synthetic spectrum}

The AIA and XRT temperature responses were used in the DEM calculation (refer section \ref{section3.6.2}).

\begin{figure}[!hbtp]
\begin{center}
\includegraphics[trim=1cm 0.5cm 1cm 0.5cm,width=0.45\textwidth]{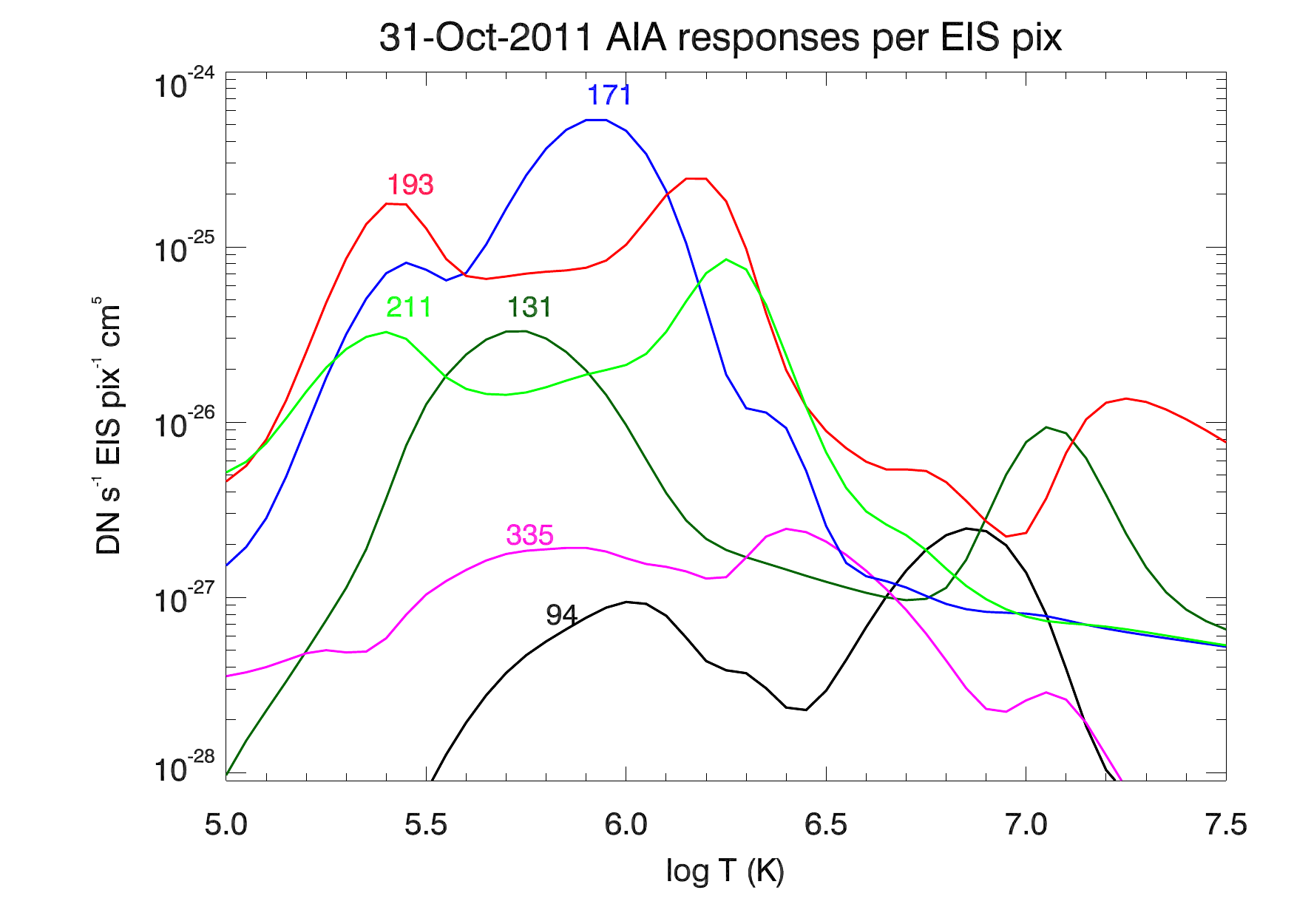}
\caption{AIA temperature responses (per EIS pixel) for the jet-spire calculated using the CHIANTI v.8 \citepads{2015A&A...582A..56D} atomic database. We used an electron density of \mbox{\textit{N}$_\textrm{e}$} = 7.6$\times$10$^{10}$ cm$^{-3}$ and photospheric abundances by \citetads{2009ARA&A..47..481A}. \label{figa1}}
\end{center}
\end{figure}


\begin{figure}[!hbtp]
\begin{center}
\includegraphics[trim=1cm 0.5cm 1cm 0.5cm,width=0.45\textwidth]{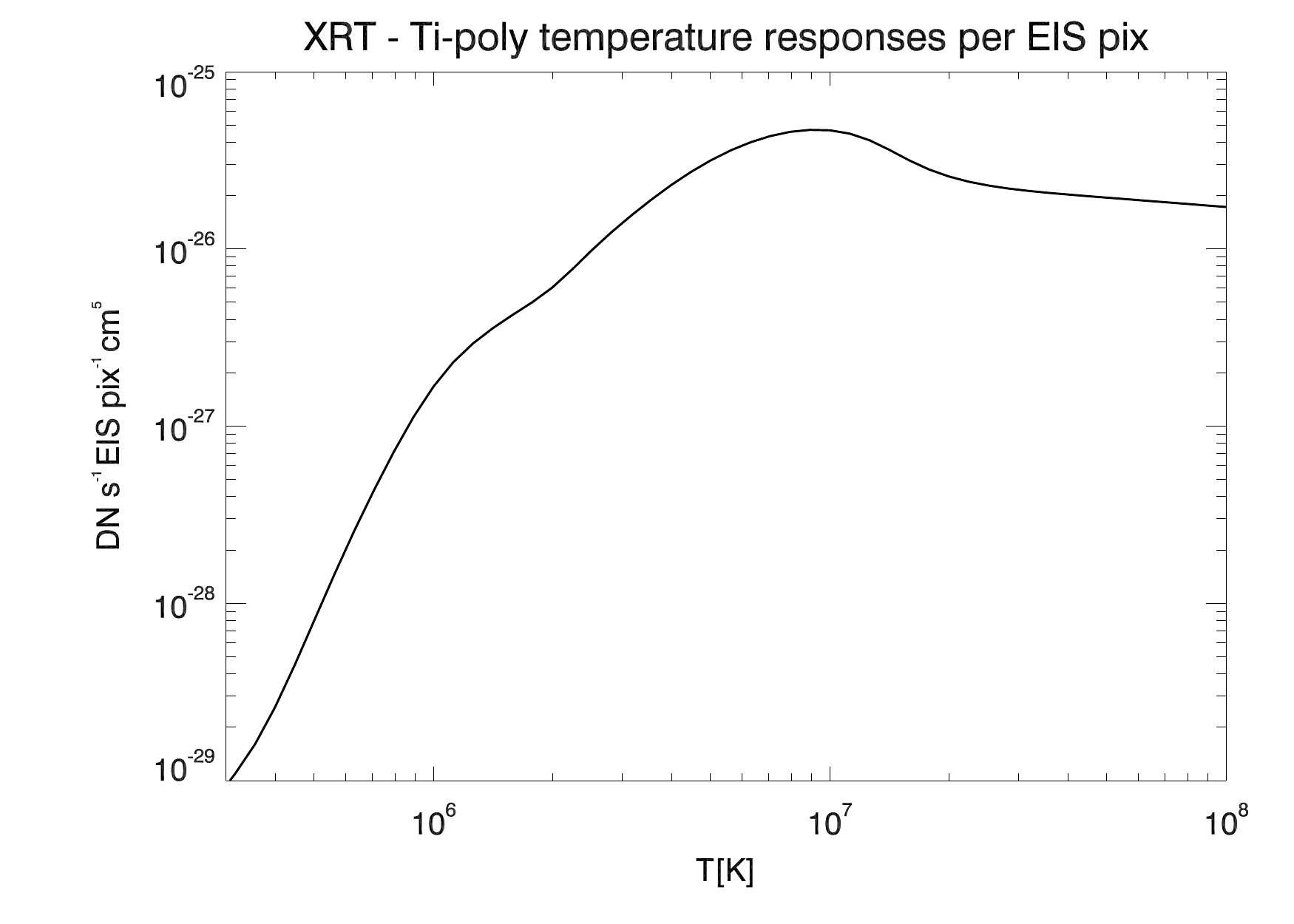}
\caption{XRT temperature responses (per EIS pixel) for the jet-spire calculated using the CHIANTI v.8 \citepads{2015A&A...582A..56D} atomic database. We used an electron density of \mbox{\textit{N}$_\textrm{e}$ = 7.6$\times$10$^{10}$} cm$^{-3}$ and photospheric abundances by \citetads{2009ARA&A..47..481A}.\label{figa2}}
\end{center}
\end{figure}


\begin{figure*}[!btp]
\begin{center}
\includegraphics[trim=0.5cm 5.5cm 0.5cm 0.2cm,width=0.73\textwidth]{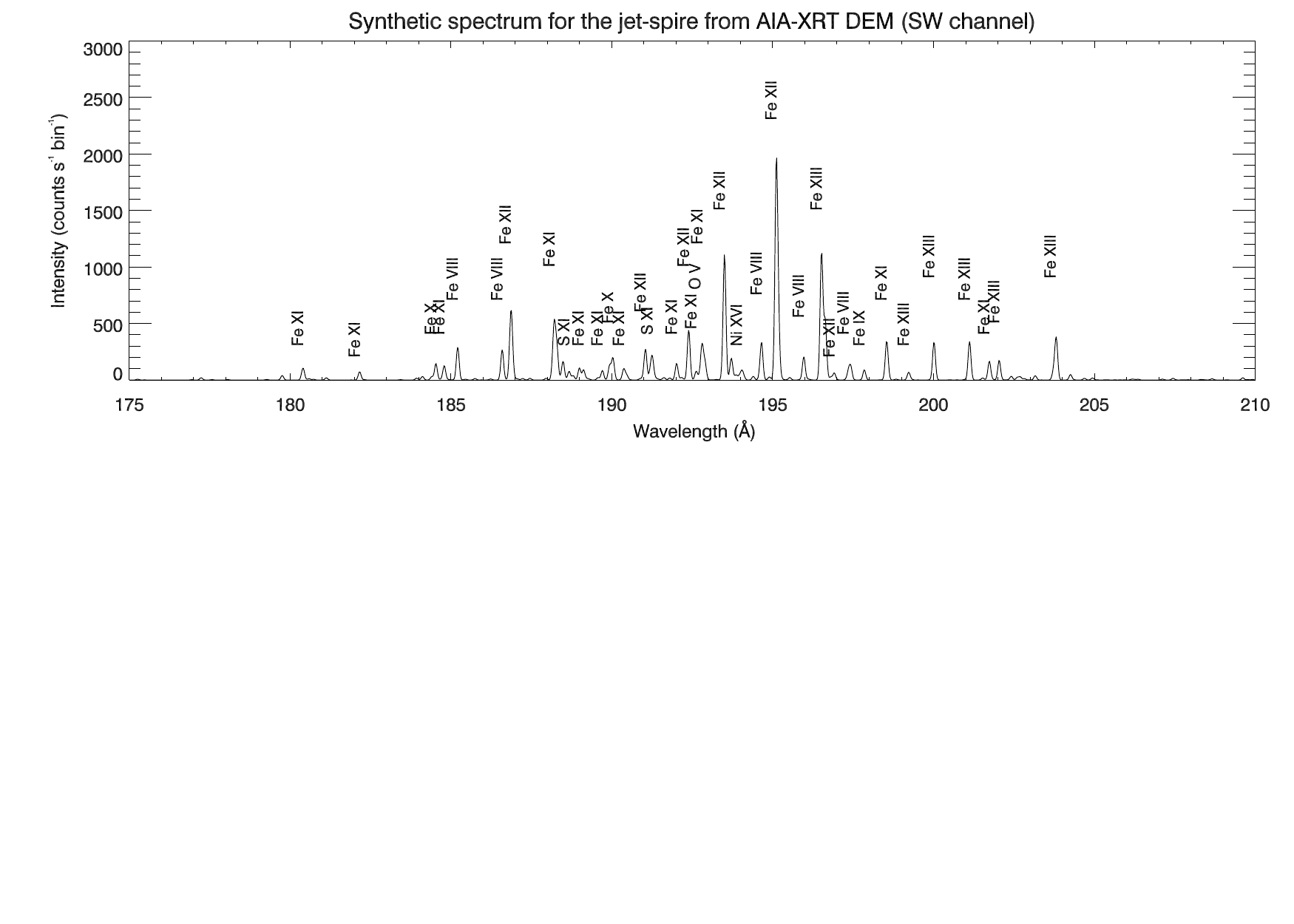}
\includegraphics[trim=0.5cm 6cm 0.5cm 0.7cm,width=0.73\textwidth]{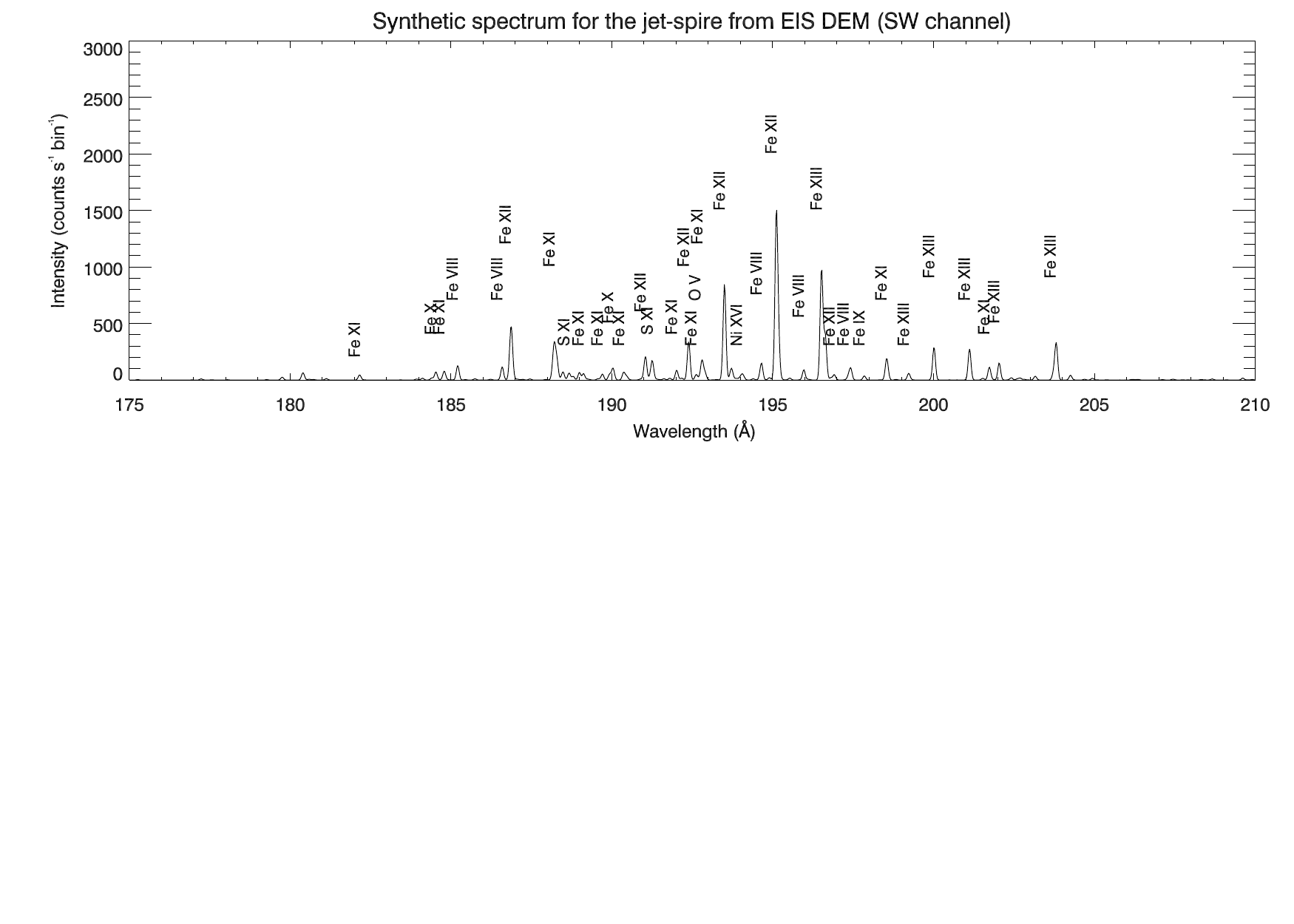}
\caption{The synthetic spectra in short wavelength channel for the jet-spire obtained from AIA-XRT DEM curves (top panel) and the EIS DEM curves (bottom panel).  \label{figa3}}
\end{center}
\end{figure*}

\begin{figure*}[!btp]
\begin{center}
\includegraphics[trim=0.5cm 5.5cm 0.5cm 0.2cm,width=0.73\textwidth]{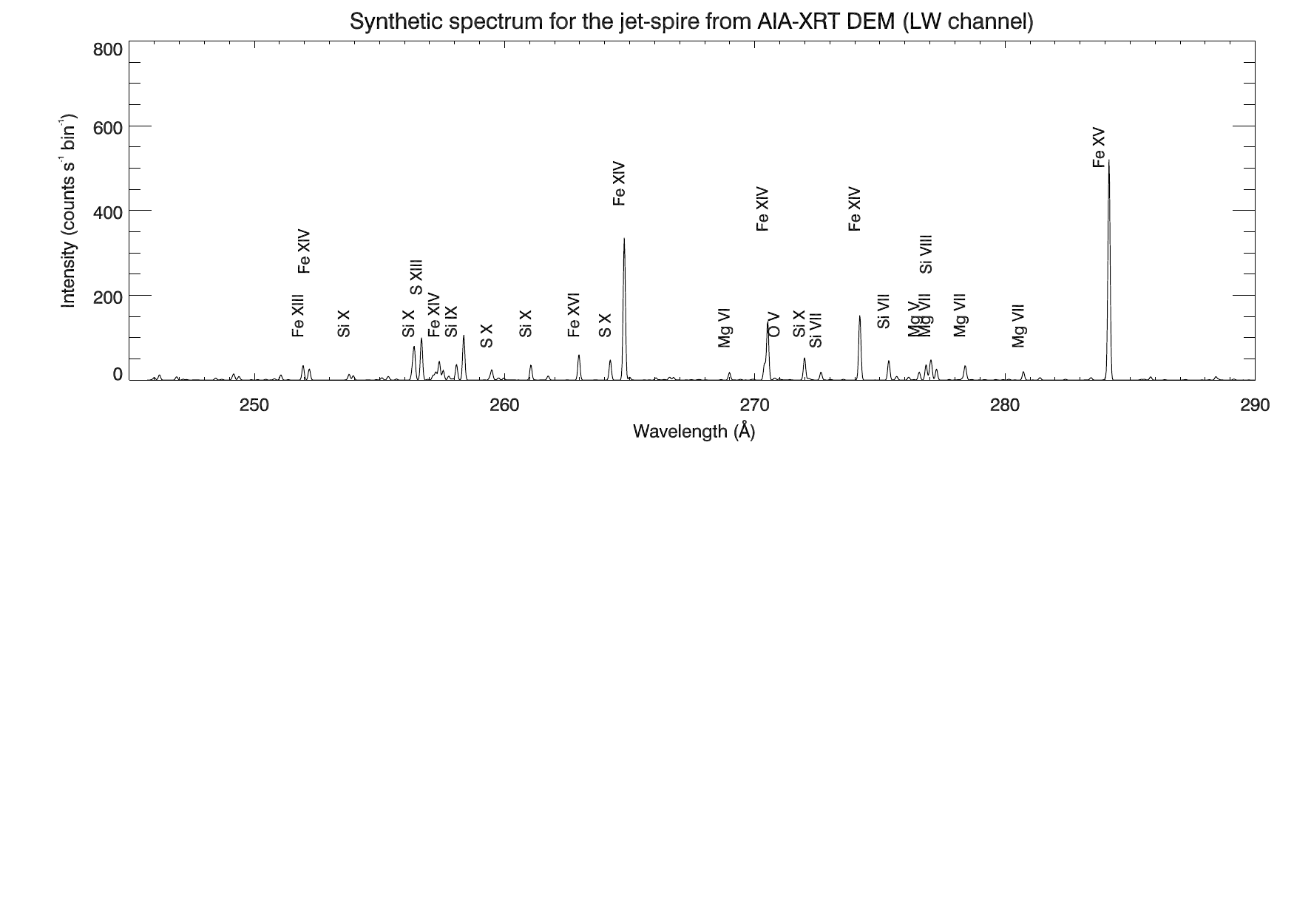}
\includegraphics[trim=0.5cm 6cm 0.5cm 0.7cm,width=0.73\textwidth]{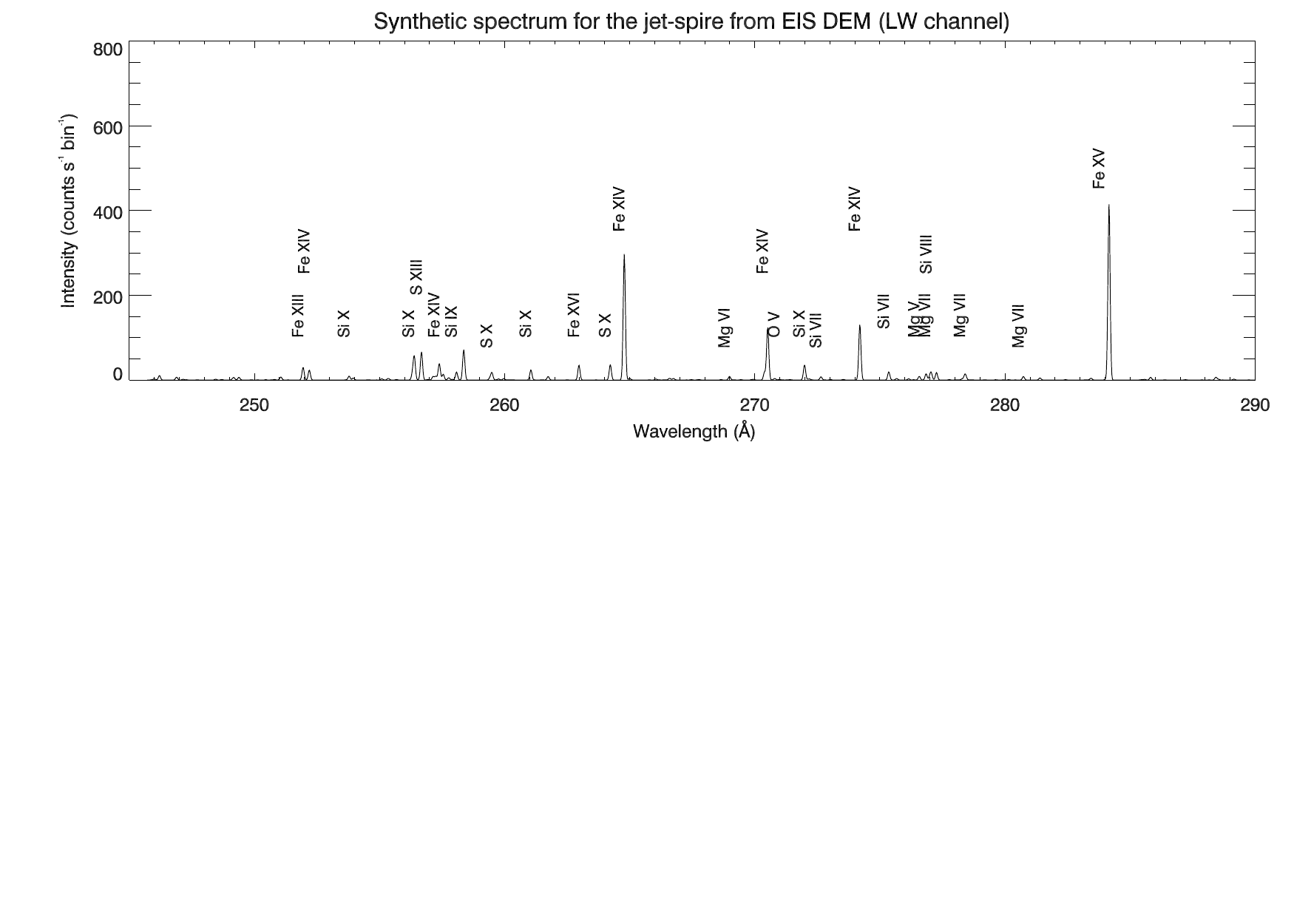}
\caption{The synthetic spectra in long wavelength channel for the jet-spire obtained from AIA-XRT DEM curves (top panel) and from the EIS DEM curves (bottom panel). \label{figa4}}
\end{center}
\end{figure*}

\begin{figure*}[!btp]
\begin{center}
\includegraphics[trim=0.5cm 5.5cm 0.5cm 0.2cm,width=0.73\textwidth]{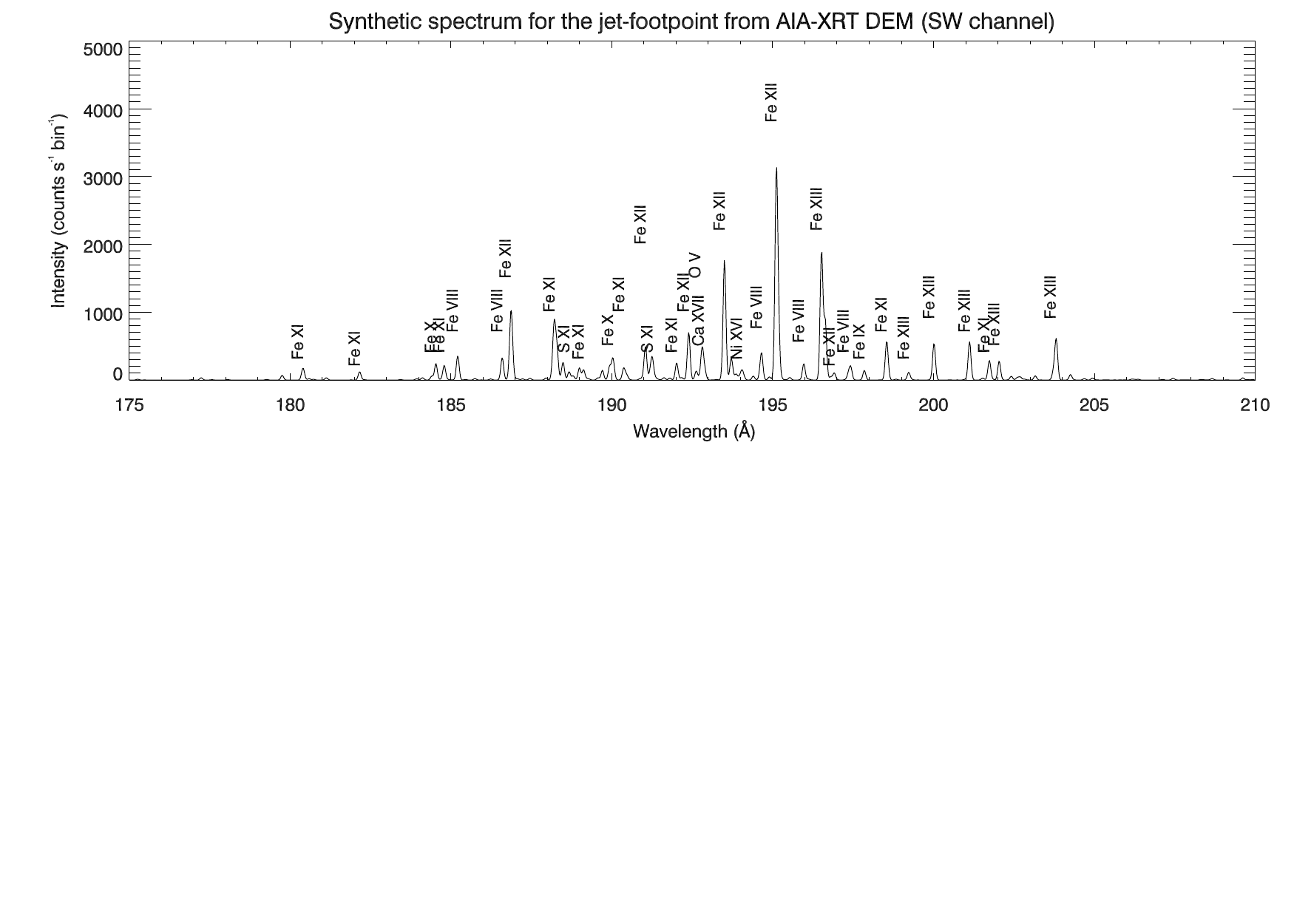}
\includegraphics[trim=0.5cm 6cm 0.5cm 0.7cm,width=0.73\textwidth]{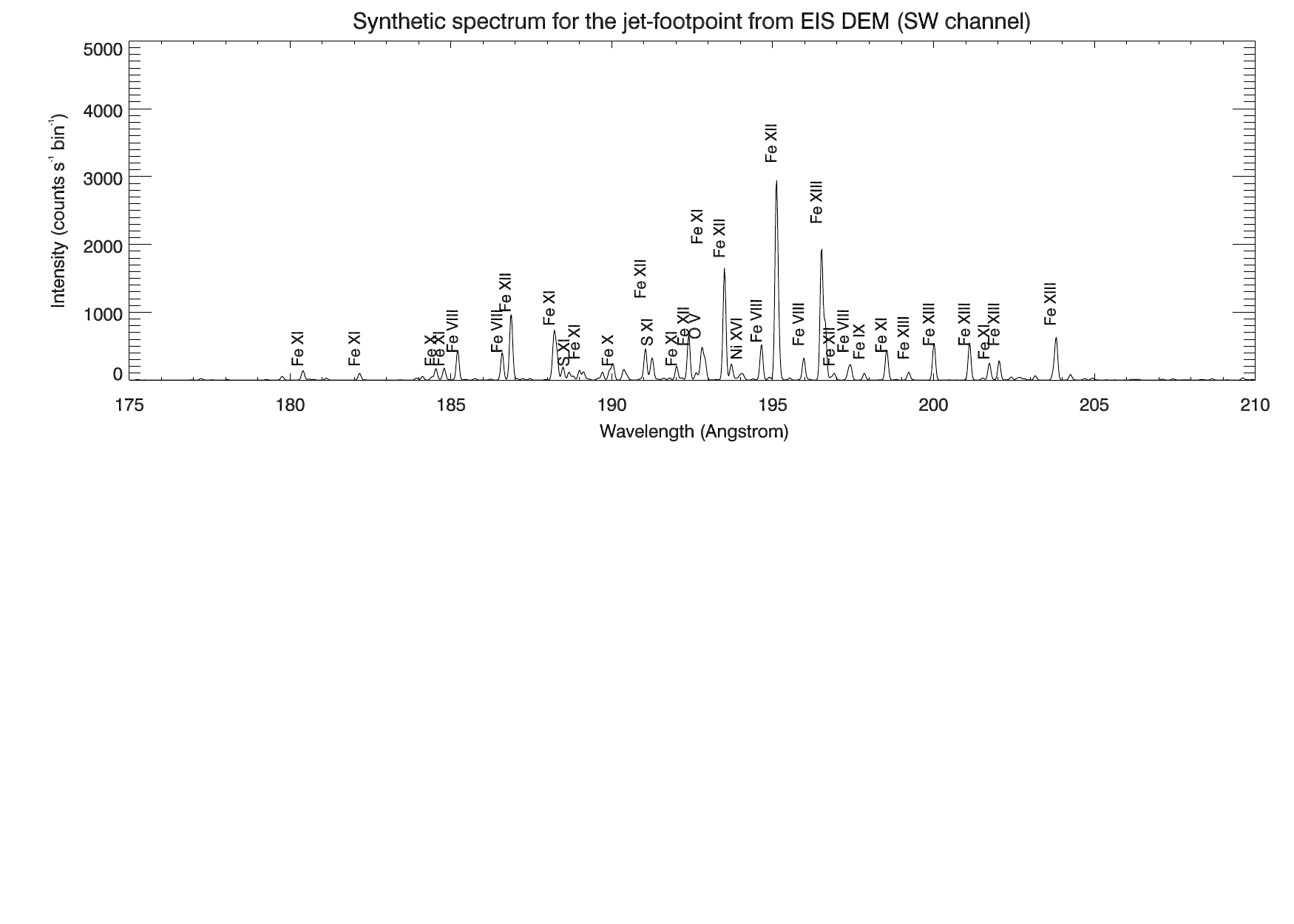}
\caption{The synthetic spectra in short wavelength channel for the jet-footpoint obtained from AIA-XRT DEM curves  and from the EIS DEM curves (bottom panel).\label{figa5}}
\end{center}
\end{figure*}

\begin{figure*}[!btp]
\begin{center}
\includegraphics[trim=0.5cm 5.5cm 0.5cm 0.2cm,width=0.73\textwidth]{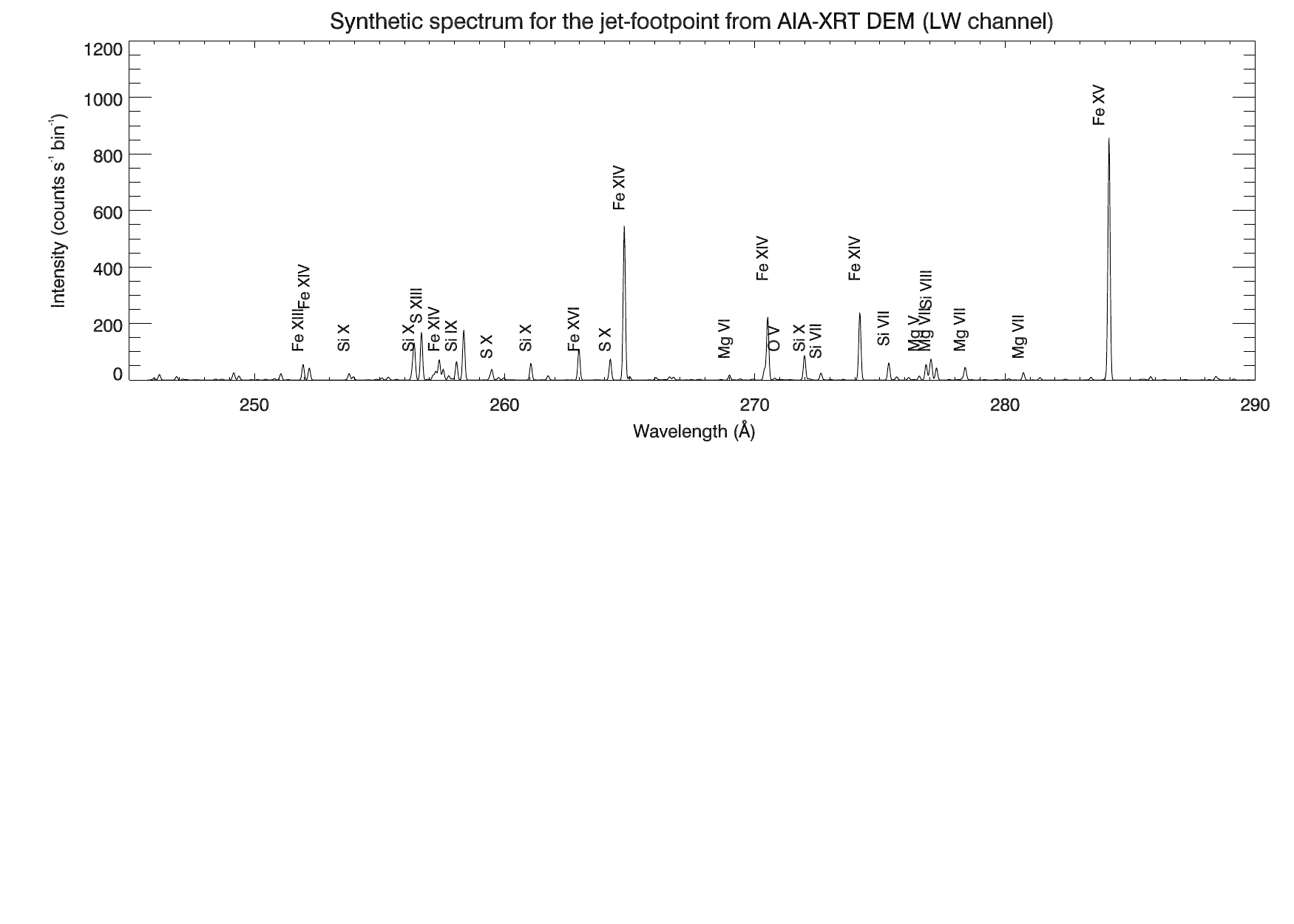}
\includegraphics[trim=0.5cm 6cm 0.5cm 0.7cm,width=0.73\textwidth]{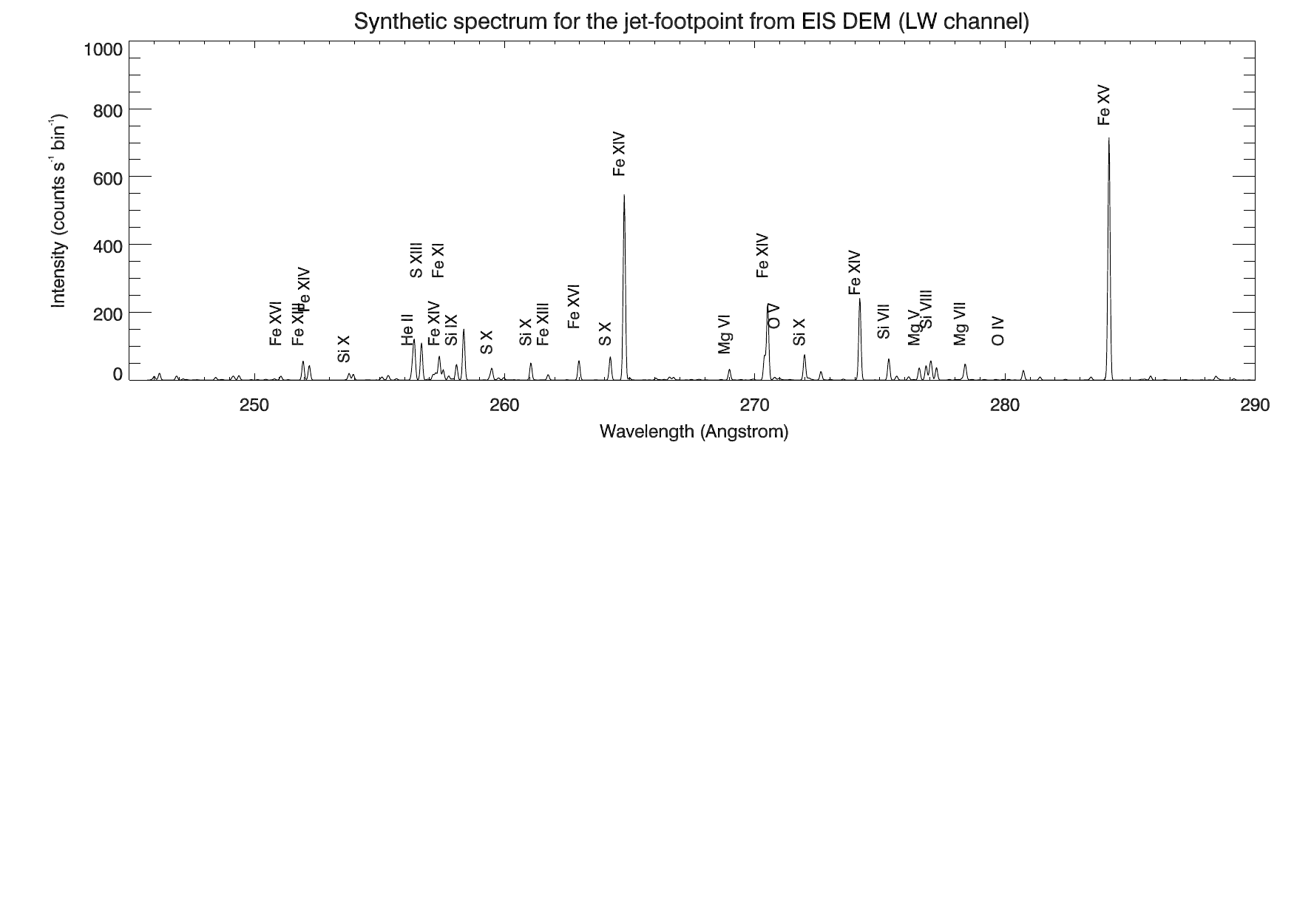}
\caption{The synthetic spectra in long wavelength channel for the jet-footpoint obtained from AIA-XRT DEM curves (top panel) and from the EIS DEM curves (bottom panel) .\label{figa6}}
\end{center}
\end{figure*}

Figures \ref{figa3} and \ref{figa4} show synthetic spectra for the SW and LW channel obtained from the EIS DEM and AIA-XRT DEM for the region of the jet-spire. Figures \ref{figa5} and \ref{figa6} show synthetic spectra for the region of the jet-footpoint.

\subsection{Cross-calibration of AIA and EIS using Moss region}

\begin{figure}[!hbtp]
\begin{center}
\includegraphics[trim=1cm 0.5cm 1cm 0.1cm,width=0.45\textwidth]{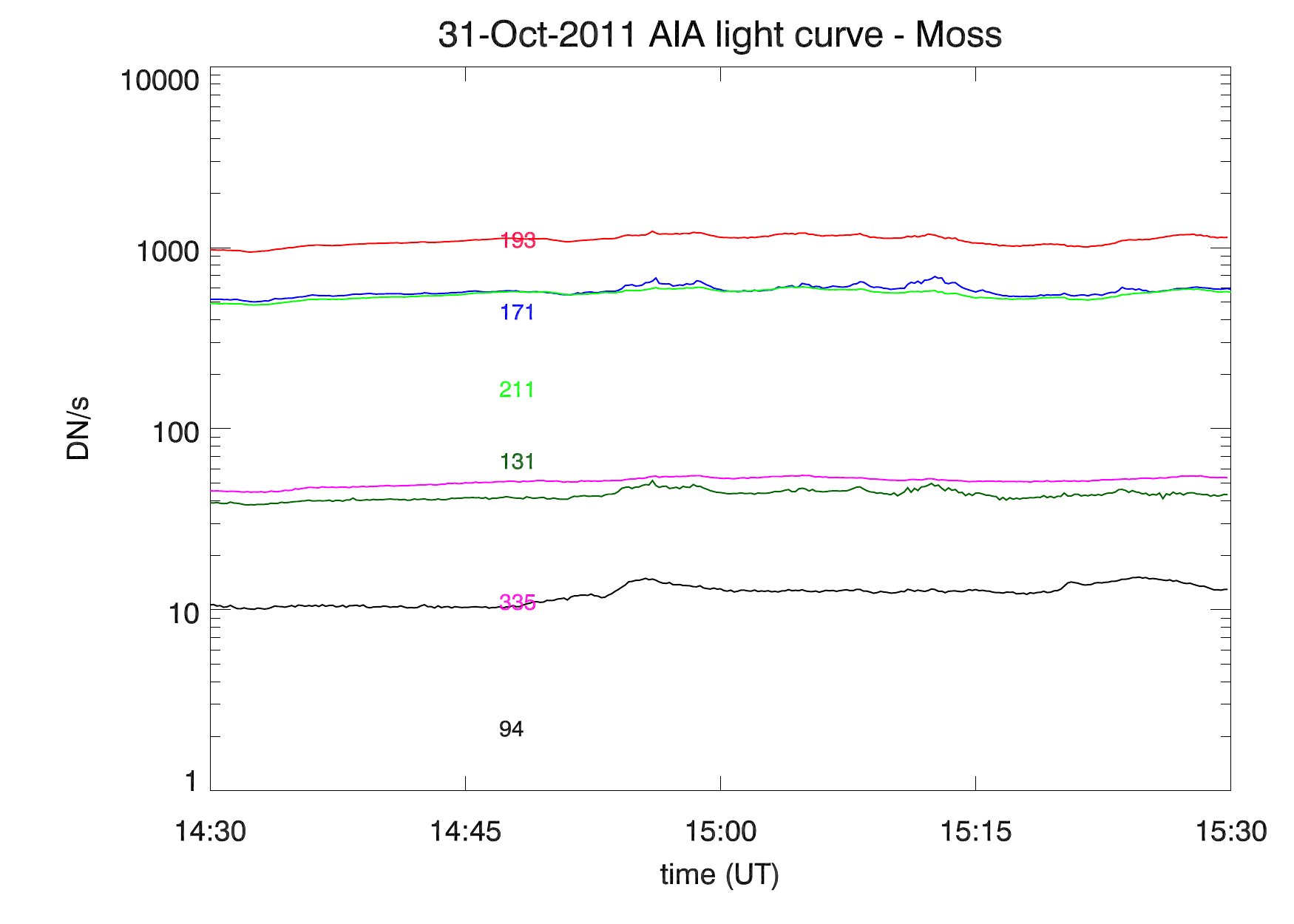}
\caption{Temporal evolution of the moss region calculated in the small yellow box shown in fig. \ref{fig1} \label{figa9}}
\end{center}
\end{figure}

\begin{figure}[!hbtp]
\begin{center}
\includegraphics[trim=1.5cm 12.0cm 0.1cm 4.0cm,width=0.45\textwidth]{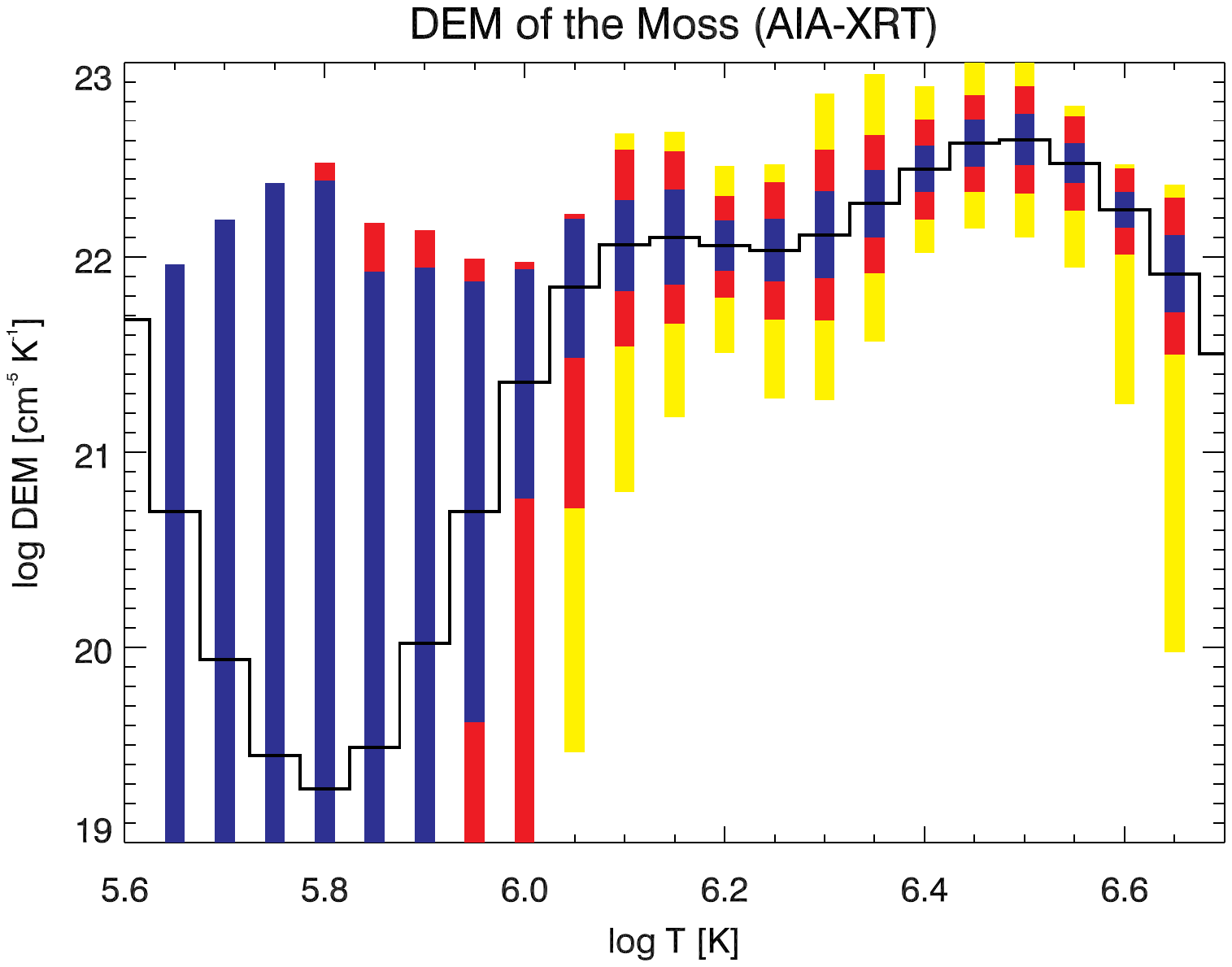}
\caption{AIA-XRT DEM curve for the moss region. The blue rectangles represents 50\%, red rectangles 80\% and yellow rectangles 95\% of the MC solutions in each temperature bin. \label{figa7}}
\end{center}
\end{figure}

\begin{figure}[!hbtp]
\begin{center}
\includegraphics[trim=1.5cm 12.0cm 0.8cm 3.0cm,width=0.45\textwidth]{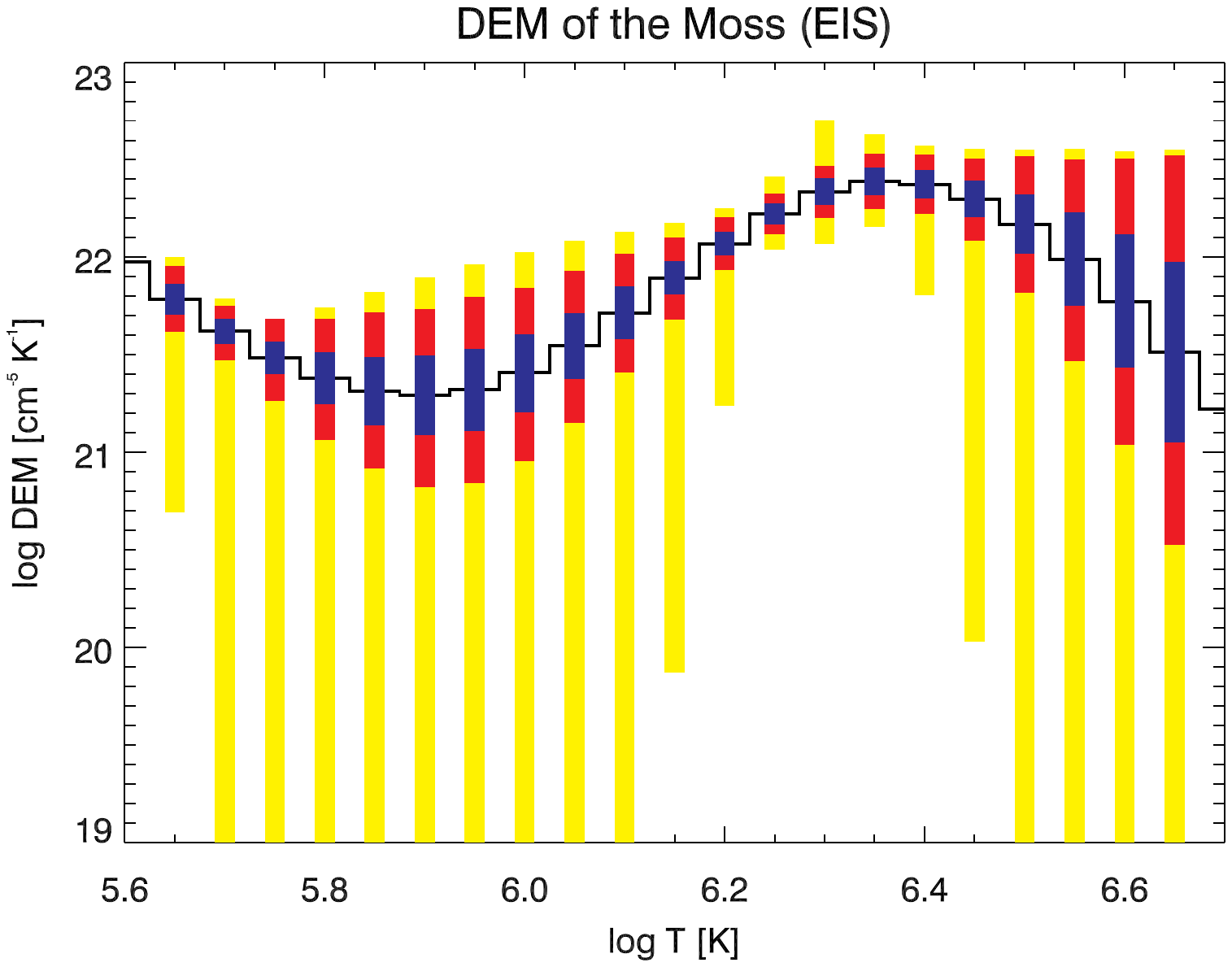}
\caption{EIS DEM curve for the moss region. The blue rectangles represents 50\%, red rectangles 80\% and yellow rectangles 95\% of the MC solutions in each temperature bin. \label{figa8}}
\end{center} 
\end{figure}

In order to investigate the cross-calibration of the AIA and EIS instruments, we selected a moss region (shown by yellow box in fig. \ref{fig1}). This region was used to verify our method of analysis (see section~\ref{section4}). We performed a similar analysis shown in section~\ref{section3.6} on the moss region. We obtained a light curve (see fig. \ref{figa9}) from all AIA channels and found that the moss region was not varying over the period of observations. The \mbox{AIA-XRT} DEM curves (see fig. \ref{figa7}) and EIS (see fig. \ref{figa8}) in the moss region were consistent over the temperature range from \mbox{log \textit{T} [K] = 5.8} to \mbox{log \textit{T} [K] = 6.6}. The total count rates in each AIA channel obtained from the EIS and AIA-XRT DEMs are also found to be in agreement (see Table \ref{tablea1}). So, we have confidence in the calibration of EIS, AIA and XRT. These results also confirm that the method we used in section \ref{section3.6} for the DEM analysis combining AIA and XRT observation is reliable.

 \begin{table}[!hbtp]
 \renewcommand\thetable{A1} 
 \centering
\caption{Observed and predicted total count rates in each AIA channel obtained from the DEM in the moss region shown as yellow box in fig. \ref{fig1}.}
 \resizebox{9cm}{!} {
 \begin{tabular}{lccccc}

 \hline
 \hline
 &		&		 &		&	\\ 	 	

 &{Band}	&{Observed} 	 &{Predicted} 	&{Predicted}	\\  

 &{({\AA})}	&	 	 &{(AIA-XRT DEM)}&{(EIS DEM)}	\\  		
  
  &		&	 	 &		&	\\  		

 \hline		
 &		& 		 &		&	\\ 	 	
				  
 &{AIA 94}	&34	 	 &28 (-18\%)	&28 (-18\%)	\\	 	 	
 
 &{AIA 131}	&122	 	 &45 (-63\%)	&66 (-46\%)	\\         	

 &{AIA 171} 	&1629		 &1518 (-7\%)	&1522 (-7\%)	\\	  	 	

 &{AIA 193}	&3162		 &3020 (-5\%)	&2636 (-17\%)	\\   	 	 	
 
 &{AIA 211} 	&1574	 	 &1495 (-5\%)	&1378 (-13\%)	\\	 	 	

 &{AIA 335}	&146		 &137 (-6\%)	&102 (-30\%)	\\  	 		
  
 &		&		 &		&	\\

 \hline
 \hline

  \end{tabular} 
}
\tablefoot{Column 2 indicates the observed AIA count rates (averaged DN/s per EIS pixel). Column 3 shows the predicted total count rates, obtained from the AIA-XRT DEM, Column 4 indicates the predicted total count rates from the EIS DEM. Values in parenthesis in columns 3 and 4 show the percentage differences between observed and predicted intensities.} \label{tablea1}

\end{table}

\end{appendix}

\end{document}